\documentclass[10pt,journal,compsoc]{IEEEtran}

\usepackage[pdftex]{graphicx}
\usepackage{amsmath,amssymb}

\usepackage[caption=false,font=footnotesize,labelfont=sf,textfont=sf]{subfig}
\usepackage{booktabs}
\usepackage{tabularx}
\usepackage{threeparttable}
\usepackage{url}
\usepackage{enumitem}
\usepackage{multirow}
\usepackage{acronym}

\usepackage{multirow}
\usepackage{rotating}
\usepackage[table,xcdraw]{xcolor}

\usepackage{fancyhdr}
\newcommand\MYhyperrefoptions{bookmarks=true,bookmarksnumbered=true,
pdfpagemode={UseOutlines},plainpages=false,pdfpagelabels=true,
colorlinks=true,linkcolor={black},citecolor={black},urlcolor={black},
pdftitle={Empathy Detection from Text, Audiovisual, Audio or Physiological Signals: A Systematic Review of Task Formulations and Machine Learning Methods},
pdfsubject={systematic review article},
pdfauthor={Md Rakibul Hasan, Md Zakir Hossain, Shreya Ghosh, Aneesh Krishna, Tom Gedeon},
pdfkeywords={Empathy, Empathy Computing, Deep Learning, Machine Learning, Pattern Recognition, Systematic Review}}

\usepackage[\MYhyperrefoptions,pdftex]{hyperref}

\usepackage[backend=biber,style=ieee,natbib=true,maxcitenames=2,mincitenames=1]{biblatex}

\begin{document}

\title{Empathy Detection from Text, Audiovisual, Audio or Physiological Signals: A Systematic Review of Task Formulations and Machine Learning Methods}
%
%
%

\author{Md~Rakibul~Hasan,~\IEEEmembership{Graduate~Student~Member,~IEEE,}
        Md~Zakir~Hossain,~\IEEEmembership{Member,~IEEE,}
        Shreya~Ghosh,
        Aneesh~Krishna,
        and~Tom~Gedeon,~\IEEEmembership{Senior~Member,~IEEE}
\IEEEcompsocitemizethanks{\IEEEcompsocthanksitem All authors are with the School of Electrical Engineering, Computing and Mathematical Sciences, Curtin University, Perth WA 6102, Australia.\protect\\
E-mail: \{Rakibul.Hasan, Zakir.Hossain1, Shreya.Ghosh, A.Krishna, Tom.Gedeon\}@curtin.edu.au
\IEEEcompsocthanksitem M. R. Hasan is also with BRAC University, Bangladesh.
\IEEEcompsocthanksitem M. Z. Hossain is also with The Australian National University, Australia.
\IEEEcompsocthanksitem T. Gedeon is also with the University of ÓBuda, Hungary.}
}

%



\IEEEtitleabstractindextext{%
\begin{abstract} 
Empathy indicates an individual's ability to understand others. Over the past few years, empathy has drawn attention from various disciplines, including but not limited to Affective Computing, Cognitive Science, and Psychology. Detecting empathy has potential applications in society, healthcare and education. Despite being a broad and overlapping topic, the avenue of empathy detection leveraging Machine Learning remains underexplored from a systematic literature review perspective. We collected 849 papers from 10 well-known academic databases, systematically screened them and analysed the final 82 papers. Our analyses reveal several prominent task formulations – including empathy on localised utterances or overall expressions, unidirectional or parallel empathy, and emotional contagion – in monadic, dyadic and group interactions. Empathy detection methods are summarised based on four input modalities – text, audiovisual, audio and physiological signals – thereby presenting modality-specific network architecture design protocols. We discuss challenges, research gaps and potential applications in the Affective Computing-based \textit{empathy} domain, which can facilitate new avenues of exploration. We further enlist the public availability of datasets and codes. This paper, therefore, provides a structured overview of recent advancements and remaining challenges towards developing a robust empathy detection system that could meaningfully contribute to enhancing human well-being.
\end{abstract}

}

\maketitle
\thispagestyle{fancy}
%
\IEEEpeerreviewmaketitle

\IEEEraisesectionheading{\section{Introduction}}\label{sec:introduction}
%
%
%
%

\acresetall 

\IEEEPARstart{E}{mpathy}, an involuntary and vicarious reaction to emotional signals from another individual or their circumstances \citep{hoffman1978toward}, is essential for effective communication in many aspects of human life, including social dynamics \citep{verhofstadt2016the}, healthcare \citep{jani2012role} and education \citep{aldrup2022empathy}. Empathy towards other individuals is essential for the survival of our species and contributes significantly to enhancing the quality of life and the depth of social interactions \citep{hoffman2000empathy,eisenberg2001origins}. Research on empathy has been a major topic across various disciplines, including Social Science, Psychology, Neuroscience, Health and, most recently, Computer Science \citep{hall2019empathy,paiva2017empathy}. Although contexts of empathy research vary across disciplines, all agree on its crucial role in human well-being \citep{hall2019empathy}. In Computer Science, a significant body of literature deals with operationalising empathy using \ac{ML} tools. 

\begin{figure}[!t]
    \centering
    \includegraphics[width=1\columnwidth]{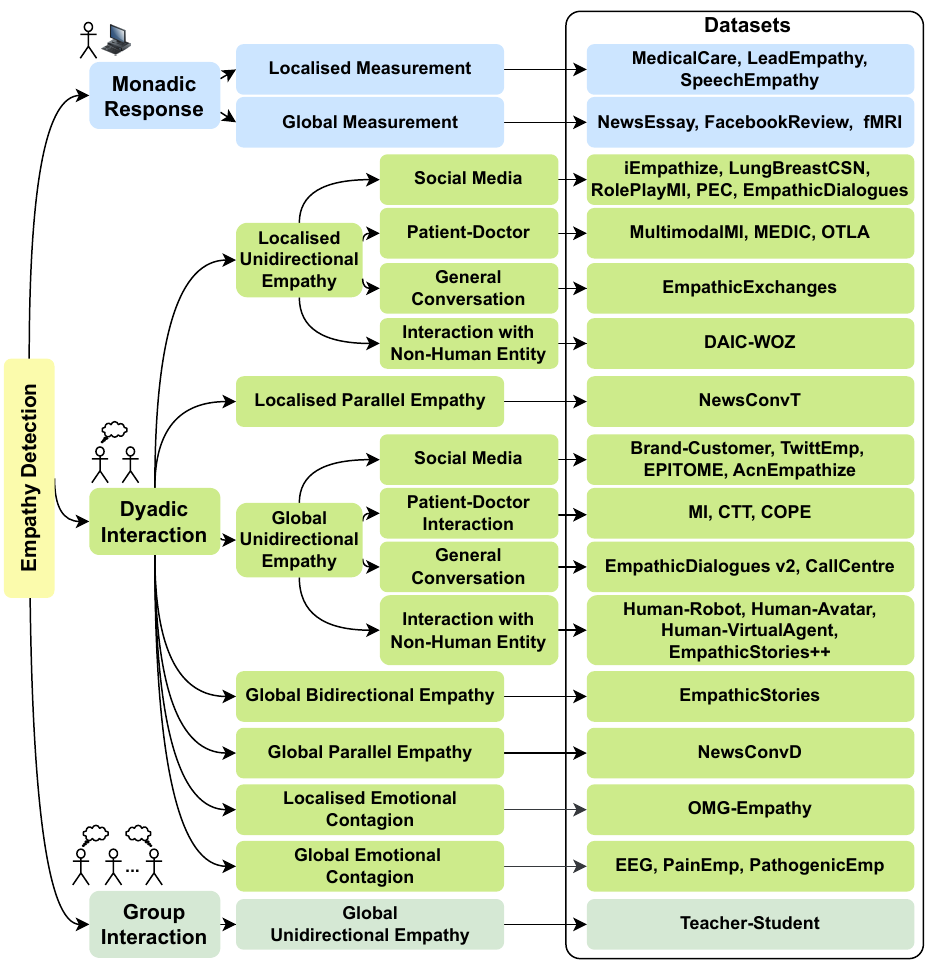}
    \caption{Hierarchy of empathy detection task formulations and representative datasets.}
    \label{fig:task_hierarchy}
\end{figure}

\ac{ML}-based empathy computing remains underexplored compared to more established domains in Affective Computing, such as emotion and facial expression recognition \citep{sariyanidi2014automatic,veltmeijer2023automatic}. Existing reviews on empathy recognition \citep{paiva2017empathy,yalcin2018computational,park2022empathy,raamkumar2022empathetic,lahnala2022critical,shetty2023scoping} are fragmented and largely limited to specific contexts. For instance, \citet{paiva2017empathy} and \citet{yalcin2018computational} focused on artificial agents (published in 2017 and 2018, respectively). In 2022, \citet{park2022empathy} reviewed empathy in social robots, and \citet{raamkumar2022empathetic} concentrated on empathic response generation in conversational systems. The more recent reviews -- \citet{lahnala2022critical} in 2022 and \citet{shetty2023scoping} in 2023 -- restricted their scope to \ac{NLP}-based approaches. In contrast, our work presents the first comprehensive \textit{systematic} literature review of empathy detection across four modalities (text, audio, video and physiological signals) and various interaction contexts (e.g., social robots, healthcare and education). We capture studies up to May 2025, and unlike prior reviews, we screen \textit{all} published works instead of cherry-picking.

\begin{figure}[!t]
    \centering
    \includegraphics[width=1\columnwidth]{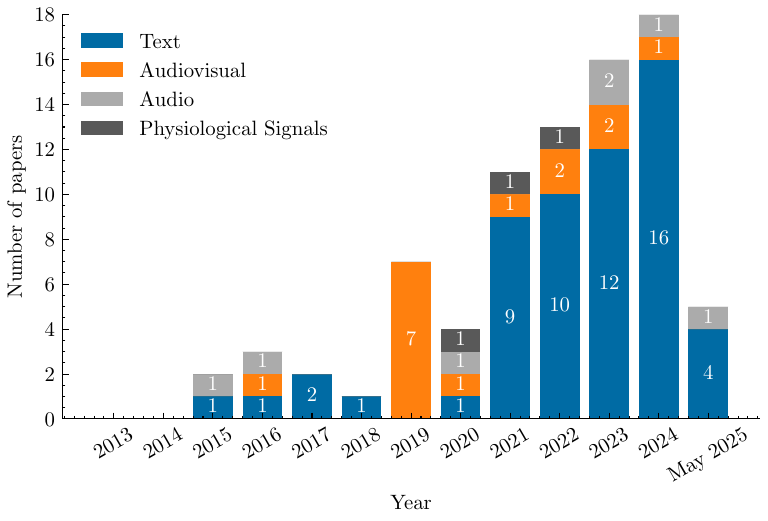}
    \caption{Growth of ML-based empathy detection literature from 2013 to 2025 (May). There are 57 text-based, 15 audiovisual-based, 7 audio-based and 3 physiological signal-based studies.}
    \label{fig:paper-vs-year}
\end{figure}

Our method adopts relevant aspects of the PRISMA 2020 guidelines \citep{page2021prisma}, which are applicable to typical systematic reviews in Affective Computing. Refer to Appendix C for a mapping of how each relevant guideline has been considered in our study. We first devised search keywords and examined ten databases, including Scopus, Web of Science and IEEE Xplore. We screened the resulting 849 records against seven inclusion and exclusion criteria. Through rigorous title-and-abstract and full-text screenings, the final 82 papers are thoroughly reviewed in this paper. As shown in \autoref{fig:paper-vs-year}, the distribution of papers reveals a predominant focus on text-based empathy detection (n = 57), followed by audiovisual (n = 15), audio (n = 7), and physiological signals (n = 3). Refer to Appendix A for the details of our paper selection strategy.

We categorise datasets based on various task formulations found in the literature (\autoref{fig:task_hierarchy}) and analyse \ac{ML} methods across four input modalities: text, audiovisual, audio and physiological signals. The overarching objective of this study is to systematically review all \ac{ML}-based empathy detection works published between 2013 and May 2025. Our main contributions are as follows:
\begin{enumerate}[label=\roman*., left=0pt]
    \item Systematic categorisation of task formulations in empathy computing.
    \item Comparative analysis of 45 existing datasets, including their composition and accessibility.
    \item Survey of modelling approaches, highlighting commonly used \ac{ML} models and code availability.
    \item Identification of high-performing methods frequently applied to benchmark datasets.
    \item Synthesis of application domains where empathy detection can be deployed.
    \item Discussion of key challenges and open opportunities across task types and modelling strategies.
\end{enumerate}

The paper is organised as follows. \autoref{sec:prelim} defines empathy and empathy detection based on several seminal works in Psychology. \autoref{sec:task-form} introduces various task formulations with a comprehensive overview of representative datasets. Our dataset analysis includes their statistics, annotation protocol and public availability of the whole annotated dataset. \autoref{sec:modality-methods} presents \ac{ML} models specific to four input modalities -- text sequences, audiovisual content, audio signals and physiological signals -- where we discuss studies involving the datasets introduced in \autoref{sec:task-form}. In presenting empathy detection works, we report public availability of the software code, best-performing models and their performance. Both \autoref{sec:task-form} and \autoref{sec:modality-methods} end with a consolidating discussion on findings, challenges and opportunities. We present some applications of empathy detection in \autoref{sec:appl} and conclude the paper in \autoref{sec:concl}.

\section{Preliminaries}\label{sec:prelim}

\subsection{Empathy and Related Concepts}
With broad usage across various disciplines, the definition of empathy varies. \citet{cuff2016empathy} reviewed 43 discrete definitions of empathy and identified eight themes related to its nature. Themes include distinguishing empathy from similar concepts and determining whether it is cognitive or affective. The term `empathy', as defined by \citet{hoffman1978toward}, is predominantly an involuntary and vicarious reaction to emotional signals from another individual or their circumstances. In another work, \citet{hoffman2000empathy} defines empathy as `an affective response more appropriate to another's situation than one's own'. Empathy is a multifaceted concept that involves perceiving, understanding and sharing the emotional thoughts of others \citep{goleman2020emotional}. It can also be defined as a multidimensional concept, such as four-dimensional empathy (perspective taking, fantasy, empathic concern and personal distress) \citep{davis1980multidimensional}, and two-dimensional empathy (empathic concern and personal distress) \citep{batson1987distress}.

Numerous endeavours have been made to disentangle empathy from other similar concepts \citep{cuff2016empathy}. Some scholars (e.g., \citep{batson1987distress,preston2002empathy}) conceptualise empathy as a comprehensive category including emotional contagion, sympathy and compassion. Empathy is defined as comprehending another's emotions through adopting their perspective; other related psychological states include compathy (feelings shared due to shared circumstances), mimpathy (copying another's emotions without personally experiencing them), sympathy (intentionally responding emotionally), transpathy (emotional contagion, where one becomes `infected' by another's emotions) and unipathy (an elevated form of emotional contagion) \citep{ickes2003everyday,becker1931some}. Despite the inherent ambiguity in defining empathy, scholars such as \citet{ickes2003everyday} and \citet{blair2005responding} advocated separating these terms. The two most related terms are empathy and sympathy, which can be described as `feeling \textit{as}' versus `feeling \textit{for}', respectively \citep{hein2008i}. Neuroscientific evidence supports the distinction between empathy and sympathy as they have distinct neural processes \citep{decety2010neurodevelopmental}. 

While distinctions in empathy are well-recognised in Psychology, empathy computing literature often overlooks these nuances during dataset collection and labelling. This is likely because the primary focus has been on detecting the \textit{presence} of empathy in any form, typically categorised into levels such as `not empathic', `somewhat empathic', and `very empathic' \citep{montiel2022explainable} or as binary classifications like `empathic' and `not empathic'. Binary labelling, in particular, is the most prevalent approach in the literature \citep{shi2021modeling,dey2023investigating,buechel2018modeling,arahim2021assessing}. Given that empathy computing literature does not always adopt a formal definition of empathy, we chose not to exclude any studies based on their operational definitions of the concept.

Within the domain of empathy, perhaps the two most common forms are cognitive empathy and emotional (also known as affective) empathy \citep{smith2006cognitive}. Understanding someone's thoughts and perspective is known as cognitive empathy, whereas vicarious sharing of emotion is known as emotional empathy \citep{smith2006cognitive}. Cognitive empathy is closely related to the theory of mind, that is, understanding another person's mental state, such as wants, beliefs and intentions \citep{blair2005responding}. In other words, cognitive empathy is `I \textit{understand} what you feel', whereas emotional empathy is `I \textit{feel} what you feel' \citep{healey2018cognitive}. Although most empathy computing studies treat empathy as a single, consolidated construct, \citet{dey2022enriching} explicitly distinguishes between cognitive and affective empathy in a patient-doctor interaction setup.

Empathy detection differs from emotion detection, although both involve analysing human responses. Emotion detection focuses on recognising an \textit{individual}'s emotional state, such as happiness or sadness \citep{picard2000affective}. In contrast, empathy detection goes into a deeper analysis of the interactions between \textit{multiple} individuals. It considers the initial emotion expressed by one person and the emotional response of the other, whose empathy is being measured. For example, if someone expresses sadness, empathy detection would analyse how the listener emotionally supports the speaker in response \citep{hosseini2021it}.

\subsection{Empathy Measurement}
In Psychology, questionnaires are widely employed to measure self-reported empathy levels \citep{davis1980multidimensional}. These instruments typically present participants with statements or scenarios, prompting them to indicate their level of agreement or emotional response. Examples of widely used empathy questionnaires include the \ac{IRI} \citep{davis1983measuring}, the Empathy Quotient \citep{baron2004empathy}, Batson's Empathy Scale \citep{batson1987distress} and the Toronto Empathy Questionnaire \citep{spreng2009toronto}.

Empathic accuracy is another method of operationalising empathy in Psychology, which assesses how accurately an individual can infer another person's thoughts and emotions. Experimentally, it is often determined by comparing one person's reported thoughts and emotions with their partner's in a dyadic interaction \citep{hall2007sources}.

In Affective Computing, computational methods are developed to objectively measure empathy levels from verbal and non-verbal cues, such as facial expressions, tone of voice and body language. To achieve this, self-reported empathy levels through psychological questionnaires often provide necessary ground truths for training \ac{ML} algorithms. Studies that use such self-reported annotations typically align with a specific definition of empathy as defined by the chosen questionnaire. For instance, \citet{batson1987distress}'s definition has been employed for measuring empathy in written essays \citep{buechel2018modeling,tafreshi2021wassa,barriere2023findings}, while the \ac{IRI} questionnaire \citep{davis1980multidimensional,davis1983measuring} has been used in studies such as \citep{wei2021effective}. However, not all studies explicitly adhere to a specific definition of empathy. For example, datasets like \texttt{RolePlayMI} \citep{wu2021towards} and \texttt{PEC} adopt heuristic labelling approaches based on the origin of the data rather than examining each sample.

\begin{table*}[!t]
    \centering
    \footnotesize
    \begin{threeparttable}
    \caption{Task Formulations and Corresponding Datasets for Empathy Detection from \textbf{Monadic Responses}.}
    \label{tab:dataset-monadic}
    \begin{tabularx}{\textwidth}{
    >{\centering\arraybackslash}p{0.1cm}
    >{\raggedright\arraybackslash}p{2.4cm}
    >{\raggedright\arraybackslash}X
    >{\raggedright\arraybackslash}p{3cm}
    >{\raggedright\arraybackslash}p{2.4cm}
    >{\centering\arraybackslash}p{0.6cm}
    >{\centering\arraybackslash}p{0.8cm}
    @{}}\toprule
    \textbf{SL} & \textbf{Name} & \textbf{Data} & \textbf{Statistics} & \textbf{Output label\tnote{a}} &\textbf{Anno.\tnote{b}} & \textbf{Public} \\ \midrule 
\multicolumn{7}{@{}l}{\texttt{\textbf{Localised Measurement}}} \\ \addlinespace
    1 & MedicalCare \citep{shi2021modeling} & Essays on simulated patient-doctor interaction & 774 essays & \{0, 1\} & T & $\times$ \\ \addlinespace
    2 & MedicalCare v2 \citep{dey2022enriching} & Re-annotation of a subset of MedicalCare dataset & 440 essays & \{Cognitive, Affective, Prosocial, None\} & T & $\times$ \\ \addlinespace
    3 & MedicalCare v3 \citep{dey2023investigating} & Re-annotation of MedicalCare v2 dataset & 440 essays & \{0, 1\} & T & $\times$ \\ \addlinespace
    4 & LeadEmpathy \citep{sedefoglu2024leadempathy} & Leaders' email to their subordinates & 770 emails, 385 participants & $\{-4, -3, \dots, 6, 7\}$ & T & \checkmark \\
    5 & SpeechEmpathy \citep{chen2024detecting} & YouTube content, such as empathy training and therapy sessions & 346 videos (total 53h) &  \{0, 1\} & T & \checkmark \\ \midrule
\multicolumn{7}{@{}l}{\texttt{\textbf{Global Measurement}}} \\ \addlinespace
    6 & NewsEssay \citep{buechel2018modeling} & Written essays in response to news articles & 403 participants, 1,860 essays, 418 articles & $[1.0, 7.0]$, \{0, 1\} & S & \checkmark \\ \addlinespace
    7 & NewsEssay v2 \citep{tafreshi2021wassa} & Extension of the NewsEssay dataset & 564 participants, 2,655 essays, 418 articles & $[1.0, 7.0]$ & S & \checkmark \\ \addlinespace
    8 & NewsEssay v3 \citep{barriere2023findings} & New essays based on a subset of articles of NewsEssay dataset & 140 participants, 1,100 essays, 100 news articles & $[1.0, 7.0]$ & S & \checkmark \\ \addlinespace
    9 & NewsEssay v4 \citep{giorgi-etal-2024-findings} & Extension of NewsEssay dataset & 192 participants, 1,146 essays, 100 news articles & $[1.0, 7.0]$ & S & \checkmark \\ \addlinespace
    10 & FacebookReview \citep{arahim2021assessing} & Comments from Facebook pages & 900 reviews, 48 hospitals & \{0, 1\} & T & $\times$ \\ \addlinespace
    11 & fMRI \citep{wei2021effective} & Resting-state fMRI data from cocaine-dependent subjects & 24 subjects & $\mathbb{R}$ & S & \checkmark \\
    \bottomrule
    \end{tabularx}
    \begin{tablenotes}
        \item[a] Output labels $[x, y]$ and $\{x, ..., y$\} refer to continuous and discreet labels, respectively. \{0, 1\} means binary labels \{No Empathy, Empathy\}. $\mathbb{R}$ -- real number, unspecified in the paper.
        \item[b] Annotation: S -- Self; T -- Third party
    \end{tablenotes}
    \end{threeparttable}
\end{table*}

\section{Task Formulations in Detecting Empathy}\label{sec:task-form}
Let $X$ be the input content on which empathy $y$ will be measured. The content can be multimodal, i.e., $X \in \{X^s, X^a, X^v\}$, where $X^s$, $X^a$, $X^v$ refer to text, audio, and video sequences, respectively. The empathy label $y$ can be formulated as different levels of empathy, as in a classification problem, or a degree of empathy, as in a regression problem. The content $X$ can consist of $N$ segments $X_i$, where $i \in [1, N]$. Accordingly, measuring empathy on some segments of the content $X_i = [x_i, x_{i+1}, \cdots, x_{i+m}]$, where $0 \leq m < N$, can be interpreted as \textit{localised} measurement, whereas measuring empathy on the whole content $X$ can be interpreted as \textit{global} measurement. 

A range of experimental setups have been proposed in the literature to define and structure the specific goals for detecting empathy. Firstly, in \textit{Monadic Response}, the focus is on measuring empathy on self-contained, individual responses of a person. Secondly, in \textit{Dyadic Interaction}, empathy is measured from the interactions between two individuals. Expanding beyond dyads, in \textit{Group Interaction}, empathy is measured on multiple individuals engaging in an interaction. Task formulations in each of these setups and corresponding datasets are summarised in the following subsections.

\subsection{Monadic Response}
Several studies explored monadic response-based empathy computing in either the localised or global manner (\autoref{tab:dataset-monadic}).

\subsubsection{Localised Measurement}
In detecting empathy at a localised level of monadic responses, \citet{shi2021modeling} proposed \texttt{MedicalCare} dataset, which consists of 774 narrative essays of simulated patient-doctor interactions written by pre-med students. Sentences of the essays were labelled as either `empathic' or `non-empathic' by six trained undergraduate students, followed by two meta-annotators. Samples were considered `empathic' if they displayed cognitive or affective empathy. As an extension of this task formulation, \citet{dey2022enriching} selected 440 essays from the pool of 774 \texttt{MedicalCare} essays and re-annotated them into four labels: cognitive empathy, affective empathy, prosocial behaviour and non-empathy, hereinafter referred to as the \texttt{MedicalCare v2} dataset. \citet{dey2022enriching}'s task formulation, therefore, aims to measure different types of empathy rather than different levels of generalised empathy like \citet{shi2021modeling}. As a further extension of the \texttt{MedicalCare v2} dataset, \citet{dey2023investigating} formulated an empathy versus non-empathy classification problem by collapsing cognitive, affective and prosocial labels into a single `empathic' class, hereinafter referred to as the \texttt{MedicalCare v3} dataset. In this updated dataset, the authors also identified four themes that a healthcare provider might express during the interactions: (1) empathic language, (2) medical procedural information, (3) both empathy and information and (4) neither empathy nor information. Such a thematic approach can help healthcare providers communicate effectively, given that they need to empathise as well as deliver procedural information. 

\texttt{LeadEmpathy} is another dataset modelling localised measurement, which consists of emails from participants acting as leaders in a business organisation \citep{sedefoglu2024leadempathy}. In the first phase of data collection, each participant wrote an email to their subordinate employees regarding a hypothetical concern that resulted in losing a customer. In the second phase of the experiment, the participants were asked to rewrite their earlier emails to increase empathy. Both emails were considered for empathy detection, and exploratory data analysis supports increased empathy in the second email. The annotation protocol considers empathy success and failure in both cognitive and affective empathy. Segments of emails are annotated into discrete empathy scores ranging from $-4$ to $+7$, which facilitates modelling it either as a classification task or a regression task. In addition, scores of 1 and below can be mapped to `low' empathy and scores of 2 and higher to `high' empathy in a binary classification setup \citep{sedefoglu2024leadempathy}.

\subsubsection{Global Measurement}
In predicting empathy on the whole content level, the \texttt{NewsEssay} dataset consists of essays written by Amazon Mechanical Turk participants about news articles involving harm to individuals, groups or nature. In addition to the essays, the dataset consists of participants' demographic information, such as age, gender, income and education level. This dataset has gone through a series of enhancements, serving empathy detection challenges in a conference workshop named \acf{WASSA}\footnote{\url{https://workshop-wassa.github.io/}}. The \ac{WASSA} 2021 and 2022 challenges use the same \texttt{NewsEssay v2} dataset \citep{tafreshi2021wassa}, whereas the \ac{WASSA} 2023 challenge release the updated \texttt{NewsEssay v3} dataset \citep{barriere2023findings}, both of which extend from the inaugural \texttt{NewsEssay} dataset \citep{buechel2018modeling} by involving new participants in the data collection experiment. The \texttt{NewsEssay v4} dataset, used in \ac{WASSA} 2024 challenge, includes samples from the \texttt{NewsEssay v3} dataset and further extends it with additional data. Empathy labels in these datasets came from the essay writers themselves as they filled in Batson's empathy and distress scale \citep{batson1987distress}. This scale includes questions related to six empathy-related emotions (sympathetic, compassionate, tender, etc.) and eight personal distress-related emotions (alarmed, upset, worried, etc.). The responses were collected on a 7-point Likert scale, where a value of one and seven means the participant is not feeling the emotion at all and is feeling the emotion extremely, respectively. After averaging the scores across questions, this dataset's ground truth degree of empathy ranges from 1 to 7 for each written essay.

People often leave reviews on products or services through online forums, including for hospitals. \citet{arahim2021assessing} collected people's reviews on the official Facebook pages of 48 public hospitals. Two hospital quality managers labelled the reviews of this \texttt{FacebookReview} dataset into `yes' or `no' empathy. This task formulation aimed to analyse the service quality of the hospitals in addition to other characteristics such as assurance, responsiveness and reliability.

Apart from individuals' expressed contents like written essays, empathy can be measured from physiological signals since these indicate individuals' internal emotional states \citep{tapus2008socially}. The \texttt{\acs{fMRI}} dataset \citep{wei2021effective} consists of resting-state \ac{fMRI} data from 24 cocaine-addicted subjects. The subjects filled in the well-known \ac{IRI} questionnaire \citep{davis1980multidimensional,davis1983measuring}, which provides continuous empathy scores for empathy assessment.

\begin{table*}[!t]
    \centering
    \footnotesize
    \begin{threeparttable}
    \caption{Task Formulations and Corresponding Datasets for Empathy Detection from \textbf{Dyadic Interactions (Localised Measurement)}.}
    \label{tab:dataset-dyadic-local}
    \begin{tabularx}{\textwidth}{
    >{\centering\arraybackslash}p{0.1cm}
    >{\raggedright\arraybackslash}p{2.2cm}
    >{\raggedright\arraybackslash}X
    >{\raggedright\arraybackslash}p{3.0cm}
    >{\raggedright\arraybackslash}p{3cm}
    >{\centering\arraybackslash}p{0.6cm}
    >{\centering\arraybackslash}p{0.8cm}
    @{}}\toprule
    \textbf{SL} & \textbf{Name} & \textbf{Data} & \textbf{Statistics} & \textbf{Output label\tnote{a}} &\textbf{Anno.\tnote{b}} & \textbf{Public} \\ \midrule
\multicolumn{7}{@{}l}{\texttt{\textbf{Localised Unidirectional Empathy}}} \\ \addlinespace
    1 & iEmpathize \citep{hosseini2021it} & Discussions from online cancer survivors network & 5,007 sentences & \{Seeking, Providing, None\} & T & \checkmark \\ \addlinespace
    2 & LungBreastCSN \citep{khanpour2017identifying} & Discussions from online cancer survivors network (lung and breast) & 2,107 messages & \{0, 1\} & T & $\times$ \\ \addlinespace
    3 & RolePlayMI\tnote{c} \citep{wu2021towards,perez-rosas2019what} & Counselling conversations from online video sharing platforms & 253 conversations & \{0, 1\} & T & \checkmark \\ \addlinespace
    4 & PEC\tnote{c} \citep{wu2021towards,zhong2020towards} & General conversations from Reddit & 355K conversations & \{0, 1\} & T & \checkmark \\ \addlinespace
    5 & EmpathicDialogues \citep{rashkin2019towards,wu2021towards} & Dyadic conversations regarding any personal situation & 810 participants and 24,850 conversations & \{0, 1\} & T & \checkmark \\ \addlinespace
    6 & MultimodalMI \citep{tavabi2023therapist,tran2023multimodal} & Real-world motivational interviewing psychotherapy sessions & 301 patients, 16 therapists and 301 sessions (each 50--60 minutes) & [0, 1], \{High, Low\} & T & $\times$ \\ \addlinespace
    7 & MEDIC \citep{zhu2023medic} & Psychological counselling & 771 video clips (total 11 hours) & \{None, Weak, Strong\} Expression & T & \checkmark \\ \addlinespace
    8 & OTLA \citep{sanjeewa2025machine} & Telephone counselling conversation from On The Line, Australia & 57 calls, 643 segments & \{High, Low\} & T & $\times$ \\ \addlinespace
    9 & DAIC-WOZ \citep{tavabi2019multimodal,gratch2014distress} & Semi-structured interviews with virtual agent & 186 participants and 2,185 conversations & \{Negative, Positive, None\} & T & $\times$ \\ \addlinespace
    10 & EmpathicExchanges \citep{montiel2024empatheticexchanges} & Conversation between two people about emotional situations & 4949 samples & \{Little to no, Somewhat, Empathic\} & T & \checkmark \\ \midrule
\multicolumn{7}{@{}l}{\texttt{\textbf{Localised Parallel Empathy}}} \\
    11 & NewsConvT \citep{barriere2023findings} & Conversation about news articles & 140 participants and 12,601 speech-turns & $[0.0, 5.0]$  & T & \checkmark \\ \addlinespace
    12 & NewsConvT v2 \citep{giorgi-etal-2024-findings} & New data similar to NewsConvT & 192 participants and 14,472 speech-turns & $[0.0, 5.0]$  & T & \checkmark \\ \bottomrule
    \end{tabularx}
    \begin{tablenotes}
        \item[a] Output labels $[x, y]$ and $\{x, ..., y$\} refer to continuous and discreet labels, respectively. \{0, 1\} means binary labels \{No Empathy, Empathy\}.
        \item[b] Annotation: S -- Self; T -- Third party
    \end{tablenotes}
    \end{threeparttable}
\end{table*}

\subsection{Dyadic Interaction}
Similar to monadic response-based empathy computing, dyadic interactions are explored both in localised and global measurements. \autoref{tab:dataset-dyadic-local} and \autoref{tab:dataset-dyad-global} categorise datasets derived from such interactions into subcategories based on their task formulations.

\subsubsection{Localised Unidirectional Empathy}
Localised measurement of unidirectional empathy (i.e., empathy flowing in one direction) has been studied across various contexts \autoref{tab:dataset-dyadic-local}, such as social media, healthcare and general conversations.

\paragraph{Social Media}
In empathising by one person towards another, several studies leverage dyadic interaction from online forums. For example, the \texttt{iEmpathize} dataset \citep{hosseini2021it} consists of discussion threads from a website named Cancer Survivors Network\footnote{\url{https://csn.cancer.org/}}. Collected from the same website, the \texttt{LungBreastCSN} dataset \citep{khanpour2017identifying} focuses on lung and breast cancer data. While the \texttt{iEmpathize} dataset aims to detect empathy \textit{direction} (seeking, providing or none), the \texttt{LungBreastCSN} dataset aims to detect the presence of empathy (empathic and non-empathic sentences).

The \texttt{RolePlayMI} dataset \citep{wu2021towards} consists of counselling conversations from video-sharing platforms, such as YouTube and Vimeo, which were originally collected in a separate study on counselling quality analysis \citep{perez-rosas2019what}. \citet{wu2021towards} later annotated this dataset into utterance-level empathic and non-empathic categories.

Being a versatile platform, Reddit is often used for analysing data across different topics of interest. The \texttt{PEC} dataset \citep{wu2021towards,zhong2020towards} consists of general conversations from three subreddits (forums dedicated to specific topics in Reddit), which are labelled heuristically. Utterances from the `Happy' and `OffMyChest' subreddits and the \texttt{EmpathicDialogues} corpus (dyadic conversations regarding any personal situation) \citep{rashkin2019towards} are considered empathic labels, whereas samples from the `CasualConversation` subreddit are considered non-empathic labels. 

\paragraph{Patient-Doctor Interaction} 
The \texttt{MultimodalMI} dataset \citep{tavabi2023therapist,tran2023multimodal} consists of two real-world motivational interviewing sessions: (Dataset 1) students assigned to MI sessions due to alcohol-related matters, and (Dataset 2) volunteering heavy drinkers aged 17--20 years. The therapists' empathy (through a sequence of texts) is annotated primarily through a Likert scale, which is also converted to binary labels (low vs high empathy). This dataset, therefore, allows empathy detection as either a classification problem \citep{tran2023multimodal} or as a regression problem \citep{tavabi2023therapist}. Annotation protocols utilise \ac{MISC} 2.5 guidelines \citep{miller2003manual} for Dataset 1 and \ac{MITI} 3.1 code \citep{moyers2010revised} for Dataset 2. The dataset consists of speech transcripts and audio, enabling it to model a multimodal empathy detection problem.

Another multimodal dataset is the \texttt{MEDIC} dataset \citep{zhu2023medic}, which consists of video, audio and text sequences of counselling case videos. It evaluates counsellors' empathy through three mechanisms: expression of experience, emotional reaction and cognitive reaction. The `expression of experience' mechanism aims to measure a client's expression to trigger empathy from a counsellor. In contrast, the `emotional reaction' and `cognitive reaction' mechanisms aim to measure the empathy of counsellors. Five trained students annotated the speech turns into none, weak and strong expressions for each mechanism.

Lastly, mental health helpline counsellors' empathy is being measured from telephone conversations in a suicide helpline service named On The Line, Australia (\texttt{OTLA}) \citep{sanjeewa2025machine}. Segments of the audio conversations are annotated by two expert mental health practitioners using the Carkhuff
and Truax Empathy scale \citep{heck1973differential}, which is later converted to binary labels (low vs high empathy) for empathy detection.

\paragraph{Interaction with Non-Human Entity}
The \texttt{DAIC-WOZ} dataset comprises semi-structured interviews between human participants and a virtual agent, which aims to measure the empathy of the virtual agent towards the human participants. The conversations are segmented into small time windows and annotated by third-party annotators into three classes: negative empathy, positive empathy and no empathy \citep{tavabi2019multimodal,gratch2014distress}. Responses such as `That sounds really hard' are considered `negative empathy', whereas no responses, expressed fillers or responses without sentiment are considered `no empathy'.

\subsubsection{Localised Parallel Empathy}
In one study covering localised empathy measurements, both persons provide empathy to someone else (parallel empathy), such as two persons conversing and empathising with some disadvantaged people. This \texttt{NewsConvT} dataset \citep{barriere2023findings} consists of dyadic conversations regarding newspaper articles featuring harm to individuals, entities or nature. Speech turns of the conversations are annotated by independent annotators on a scale of 0 to 5. It is worthwhile to note that the same news articles were used in the \texttt{NewsEssay v3} dataset, which aims to measure the empathy of individual study participants towards others through written essays. In contrast, the \texttt{NewsConvT} dataset aims to measure the empathy of two persons in the conversations.

\begin{table*}[!t]
    \centering
    \footnotesize
    \begin{threeparttable}
    \caption{Task Formulations and Corresponding Datasets for \textbf{Dyadic Interaction (Global Measurement)}.}
    \label{tab:dataset-dyad-global}
    \begin{tabularx}{\textwidth}{
    >{\centering\arraybackslash}p{0.1cm}
    >{\raggedright\arraybackslash}p{2.15cm}
    >{\raggedright\arraybackslash}X
    >{\raggedright\arraybackslash}p{3.5cm}
    >{\raggedright\arraybackslash}p{2.6cm}
    >{\centering\arraybackslash}p{0.6cm}
    >{\centering\arraybackslash}p{0.8cm}
    @{}}\toprule
    \textbf{SL} & \textbf{Name} & \textbf{Data} & \textbf{Statistics} & \textbf{Output label\tnote{a}} &\textbf{Anno.\tnote{b}} & \textbf{Public} \\ \midrule
\multicolumn{7}{@{}l}{\texttt{\textbf{Global Unidirectional Empathy}}} \\ \addlinespace
    1 & Brand-Customer \citep{singh2022linguistic} & Customer queries and brand response from Twitter & 108 brands, 667.7K customers, and 2M tweets & \{None, Weak, Strong\} & T & $\times$ \\ \addlinespace
    2 & TwittEmp \citep{hosseini2021distilling} & Cancer and 200 high-rating empathy words-related tweets & 3,000 tweets & \{Seeking, Providing, None\} & T & \checkmark \\ \addlinespace
    3 & EPITOME \citep{sharma2020computational} & Responses towards help-seeking posts in TalkLife and Reddit & 8M posts and 26M interactions & \{None, Weak, Strong\} & T & \checkmark \\ \addlinespace
    4 & EPITOME v2 \citep{hosseini2022calibrating} & EPITOME, relabelled into two classes & 8M posts and 17M interactions & \{Positive, Negative\} & T & \checkmark \\ \addlinespace
    5 & AcnEmpathize \citep{lee2024acnempathize} & Posts, quotes and replies from an online acne support forum & 12,212 samples & \{0, 1\} & T & \checkmark \\ \addlinespace
    6 & MI \citep{gibson2015predicting} & Motivational interviews between therapists and patients of drug or alcohol use & 176 therapists and 348 sessions & \{High, Low\}, $[1.0, 7.0]$ & T & $\times$ \\ \addlinespace
    7 & MI v2 \citep{gibson2016deep} & Same as the MI dataset & 348 sessions & \{High, Low\} & -- & $\times$ \\ \addlinespace
    8 & CTT \citep{xiao2015rate,baer2009agency} & Motivational interviewing sessions of drug and alcohol counselling & 200 sessions & $[1, 7]$, \{High, Low\} & T & $\times$ \\ \addlinespace
    9 & COPE \citep{chen2020automated,tulsky2011enhancing} & Conversations between cancer patient and healthcare provider(s) & 425 sessions & \{0, 1\} & T & $\times$ \\ \addlinespace
    10 & EmpathicDialogues v2 \citep{montiel2022explainable} & Samples from a dialogue generation dataset \citep{rashkin2019towards}, re-annotated into five labels & 400 conversations & \{Not, A Little, Somewhat, Empathic, Very Much\} & T & \checkmark \\ \addlinespace
    11 & CallCentre \citep{alam2016can} & Human-human conversation in call centre & 905 conversations & \{0, 1\} & T & $\times$ \\ \addlinespace
    12 & Human-Robot \citep{mathur2021modeling} & Human participants listen to six scripted stories from a robot & 46 participants and 6.9 hours audiovisual data & \{Empathy, Less Empathy\} & S & \checkmark \\ \addlinespace
    13 & Human-Avatar \citep{hervas2016learning} & Interaction between avatar and normotypical (empathic), Down syndrome and intellectual disability people (non-empathic) & 50 participants and 24,000 interactions & \{0, 1\} & O & $\times$ \\ \addlinespace
    14 & Human-VirtualAgent \citep{kroes2022empathizing} & Human participants watched a sad virtual character in virtual reality & 28 participants and 56 surveys & $[0, 20]$ & S & $\times$ \\ \addlinespace
    15 & EmpathicStories++ \citep{shen-etal-2024-empathicstories} & Humans telling personal stories and reading others' stories with AI agent & 41 participants, 269 sessions, 53h video, 5380 utterances & \{1, 2, 3, 4, 5\} & S & \checkmark \\ \midrule
\multicolumn{7}{@{}l}{\texttt{\textbf{Global Parallel Empathy}}} \\ \addlinespace
    16 & NewsConvD \citep{giorgi-etal-2024-findings} & Dialogue and perceived empathy of other person & 192 participants, 600 dialogues & \{1, 2, ..., 7\} & S & \checkmark \\ \midrule
\multicolumn{7}{@{}l}{\texttt{\textbf{Global Bidirectional Empathy}}} \\ \addlinespace
    17 & EmpathicStories \citep{shen-etal-2023-modeling} & Personal stories from social media sites, crowdsourcing and spoken narratives & 2,000 similar story pairs & \{1, 2, 3, 4\} & T & \checkmark \\
    \bottomrule
    \end{tabularx}
    \begin{tablenotes}
        \item[a] Output labels $[x, y]$ and $\{x, ..., y$\} refer to continuous and discreet labels, respectively. \{0, 1\} means binary labels \{No Empathy, Empathy\}.
        \item[b] Annotation: S -- Self; T -- Third party; O -- Other
    \end{tablenotes}
    \end{threeparttable}
\end{table*}

\subsubsection{Global Unidirectional Empathy}
Similar to the localised measurement of unidirectional empathy, global measurement has been explored in various contexts (\autoref{tab:dataset-dyad-global}), including social media and healthcare.

\paragraph{Social Media}
Several studies assess empathy globally from social media data -- primarily Twitter, Reddit and Facebook -- which are annotated by trained annotators. For example, the \texttt{Brand-Customer} dataset \citep{singh2022linguistic} consists of Twitter threads about customer service-related queries and corresponding brand responses. The authors \citep{singh2022linguistic} aimed to estimate engagement between brands and customers into three categories of empathy: none, weak and strong empathy (of brand agents).

The \texttt{TwittEmp} dataset \citep{hosseini2021distilling} consists of cancer-related tweets labelled into three categories: seeking, providing and no empathy. In a binary classification setting, the `seeking' and `providing' samples are considered positive, and the no empathy samples are considered negative.

Apart from these, various online forums facilitate consultations and mental health support. \citet{sharma2020computational} proposed a widely-recognised empathy detection framework, named \texttt{EPITOME}, which consists of three communication mechanisms: emotional reactions, interpretations and explorations. Mental health-related help-seeking posts were collected from Reddit and TalkLife (a dedicated mental health support network) and annotated into three categories -- none, weak and strong -- for each of the three mechanisms. \texttt{EPITOME} was relabelled by \citet{hosseini2022calibrating} into two classes: weak and strong communication as the positive samples, and no communication as the negative samples, hereinafter referred to as the \texttt{EPITOME v2} dataset. 

The \texttt{AcnEmpathize} dataset consists of discussions from an online acne-related forum\footnote{\url{https://www.acne.org/}}. Adopting the annotation principle of \texttt{EPITOME} \citep{sharma2020computational}, three annotators labelled each of the discussion components (posts, replies and quotes) as either `empathic' or `not empathic'. A discussion component is labelled as empathic if any part exhibits any of the three communication mechanisms (emotional reactions, interpretations and explorations) of the \texttt{EPITOME} framework.

\paragraph{Patient-Doctor Interaction}
Global measurement has been formulated in several datasets of counselling sessions between therapists and patients. For example, the motivational interviewing dataset, named \texttt{MI} \citep{gibson2015predicting}, comprises interview sessions from clinical interviews with patients of drug or alcohol use from six clinical studies. Another similar dataset (\texttt{MI v2}) \citep{gibson2016deep} also evaluates session-level empathy in motivational interviewing. The \texttt{CTT} dataset \citep{xiao2015rate,baer2009agency} includes 200 sessions between therapists and patients of drug and alcohol abuse. The annotation includes a continuous degree of empathy between 1 and 7 to model a regression problem and a low or high empathy level to model a classification problem.

The \texttt{COPE} dataset \citep{chen2020automated,tulsky2011enhancing} consists of 425 oncology encounters between cancer patients and healthcare providers. The task of this dataset is to detect empathic interactions and filter out non-empathic ones. Two trained annotators labelled this dataset into binary labels, where empathic interaction refers to when a patient expressed negative emotions and the oncologists responded empathically.

\paragraph{General Conversation}
Apart from specialised peer support communities and patient-doctor interactions, some studies aim to measure empathy in general conversations. For example, the \texttt{EmpathicDialogues} dataset -- consisting of conversations regarding any personal situation and earlier used in localised measurement \citep{rashkin2019towards,wu2021towards} -- was relabelled into five levels of empathy (\texttt{EmpathicDialogues v2}) and used for global measurement \citep{montiel2022explainable}.

The quality of support provided by call centre staff can be measured by measuring their empathy. \citet{alam2016can} proposed \texttt{CallCentre} dataset consisting of human-to-human call-centre conversation, where conversations are labelled either empathic if the session contains at least one empathic segment or non-empathic otherwise.

\paragraph{Interaction with Non-Human Entity}
Several global empathy detection datasets include interactions between humans and non-human entities, such as avatars and robots. The \texttt{Human-Robot} data collection includes a robot telling scripted stories to human participants \citep{mathur2021modeling}. Stories were told in either first-person or third-person point of view. To measure the participants' empathy towards the robot or story content, the participants answered a custom questionnaire of eight questions on a 5-point Likert scale. Thresholding based on median statistics is used to binarise the empathy scores into two labels (`empathic' and `less empathic').

The \texttt{Human-Avatar} dataset aims to assess the empathy of human participants interacting with an avatar expressing six types of emotion \citep{hervas2016learning}. Rather than self-reported annotation or third-party annotation, each interaction is labelled as empathic for normotypical participants and non-empathic for participants having social communication disorders such as Down syndrome and intellectual disability. Such a labelling approach was formulated to diagnose social communication disorders through empathy assessment.

In \texttt{Human-VirtualAgent} dataset \citep{kroes2022empathizing}, human participants watched a virtual character expressing sadness in a virtual reality environment. Participants fill in two questionnaires, including the Toronto empathy questionnaire \citep{spreng2009toronto}, to reflect how much empathy they feel towards the agents. The questionnaire responses are leveraged as self-assessed ground truth empathy scores on a scale of 0 to 20.

Lastly, the \texttt{EmpathicStories++} dataset features human participants telling personal stories to a ChatGPT-based AI agent and reading stories shared by the agent \citep{shen-etal-2024-empathicstories}. Collected over a month-long in-the-wild deployment, the dataset includes multimodal recordings (video, audio, and text) and several state and trait surveys from 41 participants. The empathy labels are based on participants’ self-reported ratings of their empathy toward each story on a sliding scale from 1 to 5.

\subsubsection{Global Parallel Empathy}
Building on the turn-level empathy annotations in the \texttt{NewsConvT} dataset \citep{barriere2023findings}, \citet{giorgi-etal-2024-findings} released \texttt{NewsConvD}, which provides dialogue-level annotations of perceived empathy. While both datasets feature conversations between two participants discussing newspaper articles, this new annotation specifically captures each participant’s perception of their partner’s empathy towards the articles.

\subsubsection{Global Bidirectional Empathy}
Bidirectional empathy can be defined as two individuals empathising with each other. \citet{shen-etal-2023-modeling} introduced \texttt{EmpathicStories}, which features bidirectional empathy in personal stories. Pairs of personal stories are labelled in terms of how two persons empathise with each other's experiences. The authors operationalise \textit{empathic similarity} in terms of three key aspects: main event, emotion and moral of the story \citep{shen-etal-2023-modeling}.

\subsubsection{Localised Emotional Contagion}
Emotional contagion -- the process by which one person's emotions and behaviours trigger similar emotions and behaviours in others -- is an element of empathy \citep{batson1987distress,preston2002empathy}. Some studies exclusively aim to measure emotional contagion, and, as such, they are discussed in this separate category of task formulation. 

One such dataset, named \texttt{OMG-Empathy} \citep{barros2019omg}, consists of audiovisual conversations with semi-scripted stories in a speaker-listener setup. Following the conversations, the listeners answered how the story impacted their emotional state in terms of a valence score between $-1$ to $+1$. Using this dataset, an empathy detection challenge\footnote{\url{https://www2.informatik.uni-hamburg.de/wtm/omgchallenges/omg_empathy_description_19.html}} was organised, and accordingly, several empathy detection models are proposed for this dataset. This dataset offers two detection protocols: a personalised protocol, which detects the valence score of each listener across all conversations, and a generalised protocol, which detects the valence score towards each story by all listeners. 

\begin{table*}[!ht]
    \centering
    \footnotesize
    \begin{threeparttable}
    \caption{Task Formulations and Corresponding Datasets for \textbf{Emotional Contagion} and \textbf{Group Interactions}.}
    \label{tab:dataset-contag-group}
    \begin{tabularx}{\textwidth}{
    >{\centering\arraybackslash}p{0.1cm}
    >{\raggedright\arraybackslash}p{2cm}
    >{\raggedright\arraybackslash}X
    >{\raggedright\arraybackslash}p{3.5cm}
    >{\raggedright\arraybackslash}p{2.6cm}
    >{\centering\arraybackslash}p{0.6cm}
    >{\centering\arraybackslash}p{0.8cm}
    @{}}\toprule
    \textbf{SL} & \textbf{Name} & \textbf{Data} & \textbf{Statistics} & \textbf{Output label\tnote{a}} &\textbf{Anno.\tnote{c}} & \textbf{Public} \\ \midrule
\multicolumn{7}{@{}l}{\texttt{\textbf{Emotional Contagion}}} \\ \addlinespace
    1 & OMG-Empathy \citep{barros2019omg} & Speaker-listener conversations based on eight semi-scripted stories & 4 speakers, 10 listeners and 80 samples (total 80h) & $[-1, +1]$ & S & \checkmark \\ \addlinespace
    2 & \acs{EEG} \citep{kuijt2020prediction} & Participants' EEG while watching an emotional video in virtual reality & 52 participants and 52 EEG samples & $[0, 96]$, \{High, Low\} & S & $\times$ \\ \addlinespace
    3 & PainEmp \citep{golbabaei2022physiological} & Participants viewing pictures of individuals with pain or no pain & 36 participants, 36 ECG and skin conductance data & \{High, Low\} & S & $\times$ \\ \addlinespace
    4 & PathogenicEmp \citep{abdul2017recognizing} & Facebook posts and responses to a questionnaire & 2,405 participants and 1.8M posts & $\mathbb{R}$ & S & $\times$ \\ \midrule
\multicolumn{7}{@{}l}{\texttt{\textbf{Group Interaction}}} \\ \addlinespace
    5 & Teacher-Student \citep{pan2022multimodal} & Class lecture from one teacher to 5-10 students & 10 teachers, 63 lectures, 338 samples & $[0.0, 10.0]$, \{Excellent, Good\} & T & $\times$ \\
    \bottomrule
    \end{tabularx}
    \begin{tablenotes}
        \item[a] Output labels $[x, y]$ and $\{x, ..., y$\} refer to continuous and discreet labels, respectively
        \item[a] $\mathbb{R}$ -- real number, unspecified in the paper
        \item[c] Annotation: S -- Self; T -- Third party
    \end{tablenotes}
    \end{threeparttable}
\end{table*}

\subsubsection{Global Emotional Contagion}
Three datasets aim to measure emotional contagion at the global level. These datasets are collected through passive dyadic interaction, where the subjects often look at some stimuli, for example, still images, video or text sequences (\autoref{tab:dataset-contag-group}).

Some studies use subjects' physiological signals during a passive interaction. For example, the \texttt{\acs{EEG}} dataset \citep{kuijt2020prediction} contains \ac{EEG} signals from 52 participants watching an emotional video (a young girl being abused as a domestic enslaved person) in virtual reality. The \ac{EEG} signals were collected from the frontal, central and occipital regions of the brain before, during and after watching the video. Before the experiment, the participants filled in the Toronto empathy questionnaire \citep{spreng2009toronto}. Although the questionnaire allows empathy annotations to range from 0 to 96, participants' responses fell within a narrower range of 49 to 86, indicating a moderate to high level of empathy reported by the participants. Using a median split, samples are also grouped into high and low classes. Both regression and classification tasks in this dataset offer empathy detection at all three times when the EEG was collected: before, during and after.

The \texttt{PainEmp} dataset \citep{golbabaei2022physiological} comprises \ac{ECG} and skin conductance data from 36 participants with different levels of autistic traits. After viewing pictures of individuals with different pain levels, the participants filled in a questionnaire regarding cognitive and affective empathy. Although it may sound a little frightening, the painful pictures (24 in total) were, in fact, collected from eight individuals going through different levels of electrical stimulation on the back of their hands. This dataset's task is to classify cognitive and affective empathy into high or low levels.

Finally, the \texttt{PathogenicEmp} dataset aims to measure emotional contagion from Facebook posts. \citet{abdul2017recognizing} defines `pathogenic empathy' as the automatic contagion of negative emotions from others, which may lead to stress and burnout. The authors argued that this negative side of empathy is risky for the health and well-being of empathic people. In their data collection, participants answered a questionnaire, which consisted of eight questions on a Likert scale, with `not at all like me' on one end and `very much like me' on the other end of the scale. The average of the responses is considered the ground truth pathogenic empathy score.

\subsection{Group Interaction}
There is only one dataset where empathy is measured from more than two persons as a group (\autoref{tab:dataset-contag-group}). This dataset, hereinafter referred to as the \texttt{Teacher-Student} dataset, consists of 63 online audiovisual lectures in a one-to-many teaching setup \citep{pan2022multimodal}. Expert annotators label each lecture session on a scale of 0 to 10 (regression task), which is then thresholded to binarise into `Excellent' and `Good' categories (classification task). The broader aim of \citet{pan2022multimodal}'s work is to evaluate teaching quality through five characteristics of a good lecture: empathy, clarity, interaction, technical management and time management.

\subsection{Discussion: Findings, Challenges and Research Gaps}
The varieties in task formulations and trends across all datasets result in several key findings and opportunities, which are discussed in the following subsections.

\subsubsection{Prospective Task Formulations}
Although unidirectional empathy is well studied in localised and global measurements (\autoref{fig:task_hierarchy}), there is a notable gap in exploring parallel empathy. We found only one study on parallel empathy in localised measurement and another in global measurement. Parallel empathy is particularly relevant in understanding collective emotional dynamics, such as in team collaborations or group therapy sessions. Investigating these scenarios could reveal insights into how empathy propagates in multi-person interactions.

Similarly, studies are scarce in group interactions, with only one study addressing the global measurement of unidirectional empathy. Group settings capture naturalistic social environments, and therefore, empathy computing in group scenarios could advance effective collaboration and social cohesion. Overall, investigating these new task formulations could significantly enrich our understanding of the evolution and change in empathy during complex social interactions.

\subsubsection{Empathy from Observer's Physiological Signals}
Physiological signals contain essential affective cues in detecting internal states of people, which are often difficult to detect in other ways, such as classifying posed smiles \citep{hossain2020using} and pretended anger \citep{chen2017are} from their real counterparts. \citet{hossain2020using}'s observer-based smile veracity detection shows that it is possible to objectively measure subjective reactions. Considering the subjective nature of empathy, accurately assessing someone's \textit{actual} (ground truth) empathy level can be challenging, making physiological signals potentially valuable.

Out of all the task formulations we review in this paper, three studies measured empathy from subjects' physiological signals, including \ac{ECG}, \ac{EEG}, \ac{fMRI} and skin conductance. Firstly, other types of physiological signals that showed effectiveness in Affective Computing, such as pupillary response \citep{chen2017are} and blood volume pulse \citep{hossain2020using}, may be experimented with for empathy detection. Secondly, instead of physiological signals from one person, we could leverage signals from all parties involved, for example, both people in a dyadic interaction. Thirdly, detecting empathy from an observer's physiological signal could be an interesting avenue of exploration. This way, physiological signals can be collected from an observer observing the interaction in which we want to measure empathy. Then, we could investigate if the signals correlate with participants' empathy. Studying observer-based physiological signals could uncover how empathy is perceived and processed by third parties, a perspective that has implications for training empathy in professionals such as counsellors, educators and healthcare workers.

\subsubsection{Annotation Protocol}
The performance of supervised \ac{ML} models is inherently tied to the quality of the labelled data. Empathy detection datasets are annotated primarily in two ways: self-annotation and third-party annotation. Self-rated annotation is a popular way to get the data labelled for Affective Computing tasks \citep{afzal2011natural}. In empathy computing, it refers to study participants filling in empathy-related questionnaires such as the \ac{IRI} \citep{davis1980multidimensional,davis1983measuring} and the Toronto \citep{spreng2009toronto} questionnaires. In contrast, third-party annotation refers to annotations by third-party trained annotators instead of the study participants from whom the data is collected. 

The choice between self-annotation and third-party annotators remains a debated topic in the literature. Of all the datasets we examine in this paper, 13 used self-annotation, and 30 used third-party annotation. \citet{buechel2018modeling} argued that self-annotation provides a more appropriate measure of empathy than third-party annotators. \citet{shi2021modeling} used \texttt{MedicalCare} dataset, annotated by trained third-party annotators, and \texttt{NewsEssay} dataset, annotated by study participants themselves. One interesting conclusion of their study is that third-party annotation could be more robust than self-rated annotation \citep{shi2021modeling}. Using ensemble methods to combine the results of multiple third-party annotators is likely to be the most robust, which should be investigated thoroughly.

The \texttt{NewsEssay v3} and \texttt{NewsConvT} datasets use the same participants but differ in annotation protocols, with self-assessment for essays and third-party annotation for conversations, respectively. As shown in \autoref{tab:study_text_regr}, studies have achieved higher empathy detection performance in the \texttt{NewsConvT} dataset than in the \texttt{NewsEssay v3} dataset, potentially suggesting that third-party annotation provides greater consistency.

Judgment varies across individuals; for example, a certain empathic interaction can be felt as `high' by someone, whereas the same can be felt as `medium' by someone else. In this case, employing multiple third-party annotators to label many of the samples separately and subsequently testing their inter-rater reliability to reach a consensus for confounding samples should be preferred. However, a study has found that third-party annotators' conscious labelling of \textit{subjective} reactions is worse than their non-conscious judgement \citep{hossain2020using}. Therefore, it can be argued that a third-party annotator may be unable to accurately assess the perceived empathy of the subject because empathy is subjective. To come to a conclusion, both self-annotation and third-party annotation, while fixing the other aspects (such as dataset and model), would be a prospective research domain to understand more about annotation and simultaneously find an appropriate annotation scheme.

\subsubsection{Public Availability of Datasets}
Data availability facilitates reproducibility, comparative studies and benchmarking research. Among the 45 empathy detection datasets reviewed, 25 are publicly available. Challenges associated with making data public include privacy and ethical considerations. Ensuring the anonymisation of sensitive information and obtaining proper consent from participants are crucial steps, supposedly for which patient data such as \texttt{MI}, \texttt{CTT}, and \texttt{COPE} are unavailable. Additionally, there may be legal and institutional restrictions that prevent sharing of certain datasets. Addressing these challenges is essential during the early stage of planning to ensure data availability. We urge the authors of the 20 non-public datasets to take active steps towards making them public.

\section{Modality-Specific Empathy Detection Methods}\label{sec:modality-methods}
Design protocols for empathy detection methods, including the choice of preprocessing techniques and specific \ac{ML} models, are primarily influenced by the input data modality. This section outlines methods based on four input modalities -- text sequences, audiovisual contents, audio signals and physiological signals -- observed across different task formulations.

\subsection{Text Sequence}\label{sec:emp-txt}
\begin{figure}[!t]
    \centering
    \includegraphics[width=1\columnwidth]{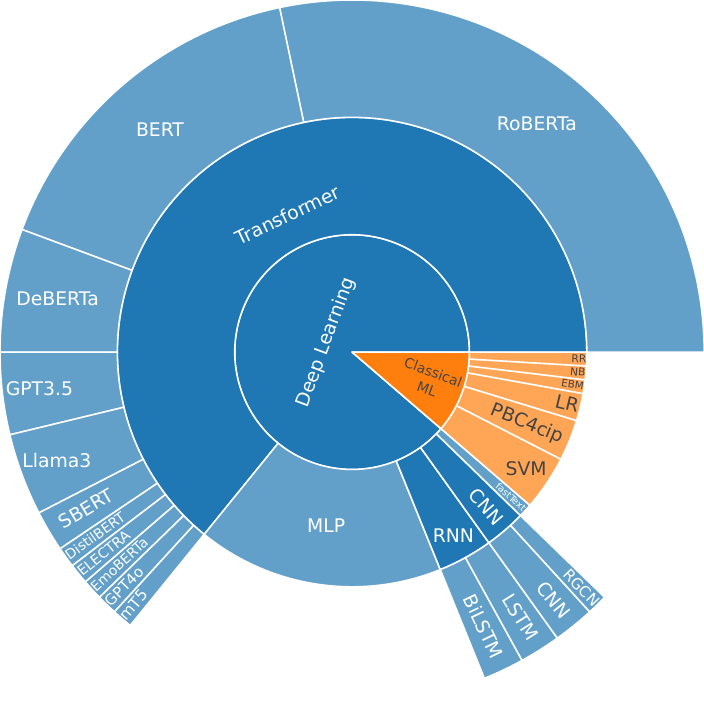}
    \caption{Usage of ML algorithms in text-based empathy detection studies, demonstrating a substantial use of pre-trained language models.}
    \label{fig:sunburst-text}
\end{figure}

In \ac{NLP} research, empathy is detected from various textual content, such as essays, conversations and social media discussions. Such text-based datasets are predominantly employed with transformer-based \ac{DL} algorithms and, to a lesser extent, with classical \ac{ML} algorithms. \autoref{fig:sunburst-text} illustrates the usage of algorithms in text-based empathy detection studies. With the recent successes of fine-tuning pre-trained language models in a variety of \ac{NLP} tasks \citep{mars2022from}, it comes as no surprise that pre-trained language models dominate the landscape of text-based empathy detection studies. Among different variants of pre-trained language models, the \ac{BERT}-based \ac{RoBERTa} is mostly used, followed by the \ac{BERT} base model itself. 

Both a continuous degree of empathy (regression) and a distinct level of empathy (classification) detection tasks are described in the literature, which are discussed in the following subsections. 

\subsubsection{Regression Task (Degree of Empathy)}
\autoref{tab:study_text_regr} summarises regression studies on common benchmark textual datasets and highlights the best-performing models in each benchmark. Refer to Appendix Table~B2 for studies on datasets used in a single published work. Most works used \texttt{NewsEssay} and its variants to detect empathy as a continuous value. The average performance in detecting empathy in essays from the \texttt{v4} and \texttt{v3} datasets is relatively lower than that observed in the \texttt{v1} and \texttt{v2} datasets, which may be attributed to the smaller size of the \texttt{v3} and the \texttt{v4} datasets. 

\begin{table}[!t]
    \centering
    \scriptsize
    \begin{threeparttable}
    \caption{Summary of Empathy Detection Studies Modelled as a \textit{Regression} Task (Degree of Empathy) on Common Benchmark Text Datasets.}
    \label{tab:study_text_regr}
    \begin{tabularx}{\columnwidth}{@{}
    >{\raggedright\arraybackslash}p{1.5cm}
    >{\centering\arraybackslash}p{0.6cm}
    >{\raggedright\arraybackslash}p{3.4cm}
    >{\raggedright\arraybackslash}p{0.6cm}
    >{\centering\arraybackslash}X
    @{}} \toprule
    \textbf{Dataset} & \textbf{Study} & \textbf{Best Model} & \textbf{PCC\tnote{b}} & \textbf{Code Avail.\tnote{a}} \\ \midrule
\multirow[t]{2}{*}{NewsEssay} & \citep{buechel2018modeling} & fastText-\acs{CNN} & 0.404 & \checkmark \\
    & \textbf{\citep{hasan2024llm-gem}} & \textbf{\acs{GPT}3.5-\acs{RoBERTa}-\acs{MLP}} & \textbf{0.924} & \checkmark \\ \midrule
\multirow[t]{2}{*}{NewsEssay v2} & \citep{butala2021team} & \acs{BERT}-\acs{MLP} & 0.473 & \checkmark \\
    & \citep{vettigli2021empna} & \acs{LR} & 0.516 & \checkmark \\
    & \citep{kulkarni2021pvg} & \acs{RoBERTa}-\acs{MLP} & 0.517 & \checkmark \\
    & \textbf{\citep{mundra2021wassa}} & \textbf{ELECTRA + \acs{RoBERTa}} & \textbf{0.558} & \checkmark \\
    & \citep{vasava2022transformer} & \acs{RoBERTa}-\acs{MLP} & 0.470 & U \\
    & \citep{ghosh2022team} & \acs{BERT}-\acs{MLP} & 0.479 & $\times$ \\
    & \citep{qian2022surrey} & \acs{RoBERTa} & 0.504 & U \\
    & \citep{lahnala2022caisa} & \acs{RoBERTa} & 0.524 & \checkmark \\
    & \citep{chen2022iucl} & \acs{RoBERTa} & 0.537 & $\times$ \\
    & \citep{plaza2022empathy} & \acs{RoBERTa} & 0.541 & $\times$ \\
    & \citep{hasan2024llm-gem} & \acs{GPT}3.5-\acs{RoBERTa}-\acs{MLP} & 0.505 & \checkmark \\ \midrule
\multirow[t]{2}{*}{NewsEssay v3} & \citep{hasan2023curtin} & \acs{BERT} & 0.187 & \checkmark \\
    & \citep{srinivas2023team} & \acs{RoBERTa}-\acs{MLP} & 0.270 & $\times$ \\
    & \citep{lu2023hit} & \acs{RoBERTa}-\acs{MLP} & 0.329 & $\times$ \\
    & \citep{wang2023ynu} & \acs{RoBERTa} & 0.331 & $\times$ \\
    & \citep{gruschka2023caisa} & \acs{RoBERTa} & 0.348 & \checkmark \\
    & \citep{chavan2023pict} & \acs{RoBERTa}-\acs{SVM} & 0.358 & $\times$ \\
    & \citep{lin2023ncuee} & \{\acs{RoBERTa}, EmoBERTa\}-\acs{MLP} & 0.415 & $\times$ \\
    & \citep{barriere2023findings} & \acs{RoBERTa} & 0.536 & $\times$ \\
    & \textbf{\citep{hasan2024llm-gem}} & \textbf{\acs{GPT}3.5-\acs{RoBERTa}-\acs{MLP}} & \textbf{0.563} & \checkmark \\ \midrule
\multirow[t]{2}{*}{NewsEssay v4} & \citep{numanoglu-etal-2024-empathify} & \{\acs{BERT}, \acs{GPT}3.5-\acs{BERT}\}-\acs{MLP} & 0.290 & $\times$ \\
    & \citep{chevi-aji-2024-daisy} & \acs{MLP} & 0.345 & $\times$ \\
    & \citep{frick-steinebach-2024-fraunhofer} & \acs{RoBERTa}-\acs{MLP} & 0.375 & $\times$ \\
    & \citep{li-etal-2024-chinchunmei} & Llama3 & 0.474 & \checkmark \\
    & \citep{kong-moon-2024-ru} & \acs{GPT}3.5 & 0.523 & $\times$ \\
    & \citep{giorgi-etal-2024-findings} & \acs{RoBERTa} & 0.629 & $\times$ \\
    & \textbf{\citep{hasan2025labels}} & \textbf{Llama3-\acs{RoBERTa}} & \textbf{0.648} & \checkmark \\ \midrule
\multirow[t]{2}{*}{NewsConvT} & \citep{hasan2023curtin} & \acs{BERT} & 0.573 & \checkmark \\
    & \citep{gruschka2023caisa} & \acs{RoBERTa} & 0.652 & \checkmark \\
    & \citep{barriere2023findings} & \acs{RoBERTa} & 0.660 & $\times$ \\
    & \citep{srinivas2023team} & \acs{RoBERTa}-\acs{MLP} & 0.665 & $\times$ \\
    & \citep{lin2023ncuee} & \{\acs{RoBERTa}, EmoBERTa\}-\acs{MLP} & 0.669 & $\times$ \\
    & \citep{wang2023ynu} & \acs{DeBERTa} & 0.674 & $\times$ \\
    & \textbf{\citep{lu2023hit}} & \textbf{\acs{DeBERTa}-\acs{MLP}} & \textbf{0.708} & $\times$ \\ \midrule
\multirow[t]{2}{*}{NewsConvT v2} & \citep{frick-steinebach-2024-fraunhofer} & \acs{RoBERTa}-\acs{MLP} & 0.034 & $\times$ \\
    & \citep{churina-etal-2024-last-min-submission} & \acs{GPT}3.5 & 0.534 & $\times$ \\
    & \citep{furniturewala-jaidka-2024-empaths} & \acs{GPT}4o-\acs{DeBERTa} & 0.534 & $\times$ \\
    & \citep{numanoglu-etal-2024-empathify} & \{\acs{BERT}, \acs{BERT}\}-\acs{MLP} & 0.541 & $\times$ \\
    & \citep{yang-etal-2024-hyy33} & \acs{DeBERTa} & 0.544 & \checkmark \\
    & \citep{lee-etal-2024-empatheticfig} & mT5-\acs{DeBERTa}-\acs{MLP} & 0.559 & $\times$ \\
    & \citep{huang-liang-2024-zhenmei} & \{\acs{BERT}, \acs{RoBERTa}, \acs{DeBERTa}\} & 0.561 & \checkmark \\
    & \citep{pereira-etal-2024-context} & \acs{RoBERTa} & 0.577 & \checkmark \\
    & \citep{li-etal-2024-chinchunmei} & Llama3 & 0.582 & \checkmark \\
    & \textbf{\citep{giorgi-etal-2024-findings}} & \textbf{\acs{RoBERTa}} & \textbf{0.694} & $\times$ \\ \midrule
    NewsConvD & \citep{li-etal-2024-chinchunmei} & Llama3 & 0.172 & \checkmark \\
        & \citep{pereira-etal-2024-context} & \acs{RoBERTa} & 0.191 & \checkmark \\
        & \citep{frick-steinebach-2024-fraunhofer} & \acs{RoBERTa}-\acs{MLP} & 0.193 & $\times$ \\
    \bottomrule
    \end{tabularx}
    \begin{tablenotes}
        \item Note: Studies using common datasets are sorted year-wise chronologically, followed by performance, where best results and methods are \textbf{bolded}.
        \item[a] U -- Unofficially available on the Internet but not provided with the paper
        \item[b] PCC -- Pearson correlation coefficient
    \end{tablenotes}
    \end{threeparttable}
\end{table}

Out of 30 studies on \texttt{NewsEssay} datasets, 28 of them leverage pre-trained word-embedding or language models. \citet{buechel2018modeling} leveraged fastText \citep{bojanowski2017enriching}, a pre-trained word-embedding model, for text embeddings, followed by a \ac{CNN} regression model, achieving a Pearson correlation coefficient of 0.404 in the \texttt{NewsEssay} dataset. \citet{vettigli2021empna} employed \ac{LR} classical \ac{ML} method on the \texttt{v2} dataset and reported a Pearson correlation coefficient of 0.516. This performance is competitive with studies utilising transformer-based language models such as \ac{BERT} and \ac{RoBERTa}, where the Pearson correlation coefficient ranges from 0.470 to 0.558 \citep{vasava2022transformer,mundra2021wassa}. This exceptional performance using classical \ac{ML} can be attributed to incorporating handcrafted features, such as lexicon-based, n-gram and demographic-based features \citep{vettigli2021empna}. Handcrafted features combined with additional raw data could be experimented with transformer architectures, as this might yield even better performance.

Instead of traditional \ac{ML} and \ac{DL} models, some recent studies \citep{hasan2024llm-gem,kong-moon-2024-ru,numanoglu-etal-2024-empathify,li-etal-2024-chinchunmei,frick-steinebach-2024-fraunhofer} leverage \ac{LLM}. For example, \citet{hasan2024llm-gem} introduces a system called \ac{LLM}-Guided Empathy (\ac{LLM}-GEm) that leverages \ac{GPT}-3.5 \ac{LLM} for three distinct purposes: converting numerical demographic numbers into semantically meaningful text, augmenting text sequences and rectifying label noises. Experiments on \texttt{NewsEssay v1}, \texttt{v2} and \texttt{v3} datasets demonstrate that LLM-GEm achieves state-of-the-art performance on the \texttt{v1} and \texttt{v3} datasets using a \ac{RoBERTa}-based pre-trained language model as the prediction model. On the \texttt{v2} dataset where \ac{LLM}-GEm underperformed, \citet{mundra2021wassa} reported the best result (Pearson correlation coefficient: 0.558) using an ensemble of ELECTRA and \ac{RoBERTa} models. Such a higher performance can be attributed to the ensemble of two language models (ELECTRA and \ac{RoBERTa}). Overall, \ac{RoBERTa} appears to be the best method in detecting empathy within the \texttt{NewsEssay} datasets.

Performance on the \texttt{NewsConvT} dataset is higher than \texttt{NewsEssay} datasets, with 0.708 as the highest Pearson correlation coefficient using a \ac{DeBERTa} model \citep{lu2023hit}. Plausible reasons could be the annotation protocols (as discussed earlier, \texttt{NewsConvT} uses third-party annotation, which is likely to be more consistent and reduce the noise in the labels) and the size of the datasets (12,601 samples in the \texttt{NewsConvT} dataset compared to 1,100--2,655 samples in the \texttt{NewsEssay} datasets).

\subsubsection{Classification Task (Level of Empathy)}
In the case of modelling empathy as a classification task, four benchmark datasets are used in multiple studies to allow comparative analysis (\autoref{tab:study_text_class}). Refer to Appendix Table~B3 for classification studies on datasets used in a single published work. 

\begin{table}[!t]
    \centering
    \scriptsize
    \begin{threeparttable}
    \caption{Summary of Empathy Detection Studies Modelled as a \textit{Classification} Task (Level of Empathy) on Common Benchmark Text Datasets.}
    \label{tab:study_text_class}
    \begin{tabularx}{\linewidth}{@{}
    >{\raggedright\arraybackslash}p{1.2cm}
    >{\centering\arraybackslash}p{0.7cm}
    >{\raggedright\arraybackslash}p{1.7cm}
    >{\raggedright\arraybackslash}p{2.7cm}
    >{\centering\arraybackslash}X
    @{}} \toprule
    \textbf{Dataset} & \textbf{Study} & \textbf{Best Model} & \textbf{Performance\tnote{a}} & \textbf{Code Avail.} \\ \midrule
    EPITOME & \citep{sharma2020computational} & \acs{RoBERTa} & Acc $\in [79.4\%, 92.6\%]$, \textbf{F1} $\mathbf{\in [62.6\%, 74.5\%]}$ & \checkmark \\
        & \citep{lee2023empathy} & \acs{SBERT}, \acs{EBM} & \textbf{Acc} $\mathbf{\in [88.3\%, 95.3\%]}$, F1 $\in [59.5\%, 62.7\%]$ & \checkmark \\ \midrule
    iEmpathize & \citep{hosseini2021it} & \acs{BERT} & F1 $\in [78.9\%, 85.8\%]$ & $\times$ \\
        & \citep{jiang2025utterance} & \acs{BERT}, \acs{BiLSTM}, \acs{RGCN} & Acc: 77.83\% & $\times$ \\ 
        & \citep{hosseini2023feature} & \acs{RoBERTa} & Acc: $81.1\%$ & $\times$ \\ \midrule
    TwittEmp & \citep{hosseini2021distilling} & \acs{BERT}-\acs{MLP} & F1 $\in [68.6\%, 85.7\%]$ & $\times$\\
        & \citep{jiang2025utterance} & \acs{BERT}, \acs{BiLSTM}, \acs{RGCN} & Acc: 74.89\% & $\times$ \\ \midrule
    \multirow{2}{*}{\parbox{1.2cm}{Empathic- Exchanges}} & \citep{cruz2025empathetic} & \acs{PBC4cip} & CEM: 0.593 & $\times$ \\
        & \citep{montiel2024empatheticexchanges} & \acs{PBC4cip} & CEM: 0.597, F1: 0.452 & \checkmark \\
    \bottomrule
    \end{tabularx}
    \begin{tablenotes}
        \item Note: Studies using common datasets are sorted year-wise chronologically, followed by performance, where best results are \textbf{bolded}.
        \item[a] Range of performance is reported when overall classification performance is unavailable. CEM -- Closeness Evaluation Measure.
    \end{tablenotes}
    \end{threeparttable}
\end{table}

\citet{sharma2020computational} used both unsupervised learning (domain adaptive pre-training) and \ac{RoBERTa}-based supervised learning on their \texttt{EPITOME} framework. In contrast, \citet{lee2023empathy} used Micromodels \citep{lee2021micromodels} as an attempt towards explainable \ac{ML}. \citet{lee2023empathy}'s approach first calculates semantic similarity scores between the \ac{SBERT} representations of some fixed seed utterances and dataset samples. The similarity scores are then used as a feature in an \ac{EBM} model to classify empathy in each of the mechanisms of \texttt{EPITOME}. In terms of quantitative scores, \citet{lee2023empathy}'s model provides better accuracy (a maximum of 95.3\% vs 92.6\%) but less F1 score (a maximum of 62.7\% vs 74.5\%) than \citet{sharma2020computational}'s models on the \texttt{EPITOME} dataset. However, one important insight from \citet{lee2023empathy}'s study is that the current empathy detection models probably consider surface-level information rather than the whole conversation context.

In detecting empathy on the \texttt{iEmpathize} dataset, \citet{hosseini2021it,hosseini2023feature} leveraged \ac{BERT} and \ac{RoBERTa} models, respectively. Despite being the same dataset, their reported classification performances are on different evaluation metrics: a maximum F1 score of 85.9\% is reported in \citep{hosseini2021it}, and a classification accuracy of 81.1\% is reported in \citep{hosseini2023feature}. The key contribution of \citet{hosseini2023feature}'s work is a data-agnostic technique for prompt-based few-shot learning to improve the performance of pre-trained language models on empathy and emotion classification tasks, especially when training data is limited and noisy.

\subsection{Audiovisual Content}\label{sec:emp-vid}
\begin{figure}[!t]
    \centering
    \includegraphics[trim={0 2cm 0 0},clip,width=1\columnwidth]{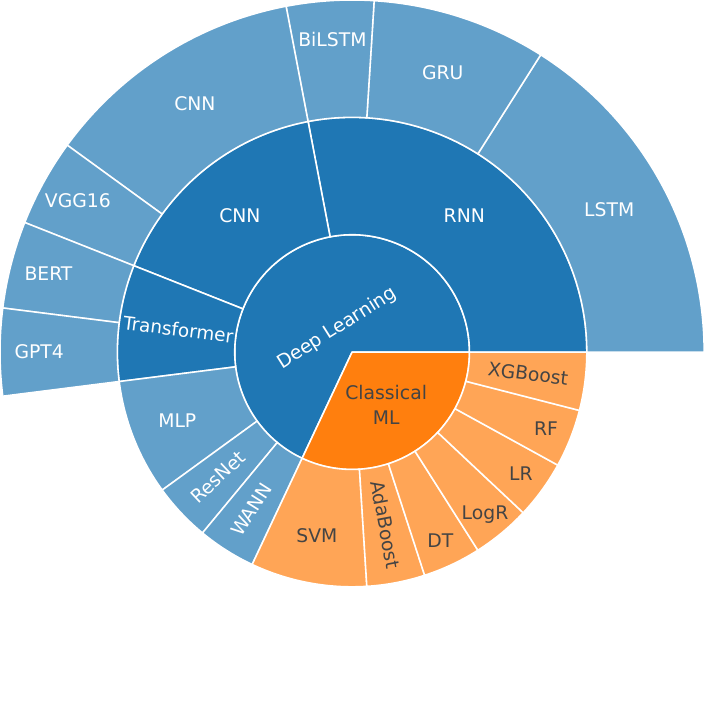}
    \caption{Usage of ML algorithms in audiovisual dataset-based empathy detection studies.}
    \label{fig:sunburst-video}
\end{figure}
Empathy detection from audiovisual contents is designed mostly as a multimodal system combining computer vision and \ac{NLP} techniques, with inputs such as facial expressions, hand gestures and audio conversations. This section, therefore, includes some multimodal approaches, which utilise audio, video and sometimes text sequences. \autoref{fig:sunburst-video} illustrates the application of algorithms in audiovisual-based empathy detection works. As usual, \ac{DL} models are the predominant choice, although classical \ac{ML} models are also widely employed. Within the \ac{DL} category, \ac{CNN} and \ac{RNN}-based models are most frequently used, whereas in the classical \ac{ML} category, \ac{SVM} enjoys a higher level of usage.

\begin{table}[!t]
    \centering
    \scriptsize
    \begin{threeparttable}
    \caption{Summary of Empathy Detection Regression Studies on the Single Common Benchmark Audiovisual Dataset: OMG-Empathy.}
    \label{tab:study_video}
    \begin{tabularx}{\linewidth}{@{}
    >{\centering\arraybackslash}p{0.7cm}
    >{\raggedright\arraybackslash}p{2.8cm}
    >{\raggedright\arraybackslash}p{2.5cm}
    >{\centering\arraybackslash}X
    @{}} \toprule
    \textbf{Study} & \textbf{Best Model} & \textbf{CCC}\tnote{a} & \textbf{Code Avail.\tnote{b}} \\ \midrule
    \citep{hinduja2019fusion} & \acs{CNN}, RF & $0.02$ (P), $0.04$ (G) & $\times$ \\
    \citep{azari2019towards} & \acs{SVM} & $0.08$ (P) & U \\
    \citep{mallol-ragolta2019performance} & \acs{BiLSTM} & $0.11$ (P), $0.06$ (G) & \checkmark \\
    \citep{tan2019multimodal} & LSTM & $0.14$ (P, G) & \checkmark \\
    \citep{barbieri2019towards} & \acs{GRU}, \acs{LSTM}, \acs{CNN}, \acs{MLP} & $0.17$ (P, G) & \checkmark \\
    \textbf{\citep{barros2019omg}} & \textbf{VGG16-\acs{LSTM}-\acs{SVM}} & \textbf{CCC $\textbf{0.17}$ (P), $\textbf{0.23}$ (G)} & $\times$ \\
    \citep{lusquino2020weightless} & \acs{WANN} & $0.25$ (Validation set) & $\times$ \\
    \bottomrule
    \end{tabularx}
    \begin{tablenotes}
        \item Note: Studies using common datasets are sorted year-wise chronologically, followed by performance, where best results and methods are \textbf{bolded}.
        \item[a] Performance refers to test-set performance unless otherwise stated. P -- Personalised protocol; G -- Generalised protocol; CCC -- Concordance Correlation Coefficient
        \item[b] U -- Unofficially available on the Internet but not provided with the paper
    \end{tablenotes}
    \end{threeparttable}
\end{table}

\autoref{tab:study_video} summarises empathy detection studies on the only common benchmark audiovisual dataset, named \texttt{OMG-Empathy}. Refer to Appendix Table~B4 for studies on audiovisual datasets used in a single published work.

\subsubsection{Regression Task (Degree of Empathy)}
The baseline model in the \texttt{OMG-Empathy} challenge \citep{barros2019omg} used VGG16 architecture to process facial expression and \ac{LSTM} to process spatial-temporal features. The outputs of these two networks were then concatenated and fed to an \ac{SVM} for empathy detection, which resulted in 0.17 and 0.23 correlation coefficients in the personalised and generalised empathy protocols, respectively. Challenge participants \citep{barbieri2019towards,hinduja2019fusion,tan2019multimodal,mallol-ragolta2019performance,azari2019towards} did not surpass these baseline results. Among them, \citet{barbieri2019towards} performed the best, achieving a 0.17 correlation coefficient for both protocols. They used separate models for each modality -- \ac{GRU} on audio signals, \ac{LSTM} on audio transcripts (text) and \ac{CNN} on vision (face and body images) -- followed by \acp{MLP}. To integrate the predictions across different modalities, they used a weighted average proportional to the validation score on each modality, followed by a Butterworth low-pass filter.

\citet{tan2019multimodal} extracted multimodal features using VGG-Face \citep{parkhi2015deep} on faces, openSMILE \citep{eyben2013recent} on audio and GloVe embedding \citep{pennington2014glove} on texts. Using a multimodal \ac{LSTM} model, they reported a correlation coefficient of 0.14 on both personalised and generalised protocols. \citet{mallol-ragolta2019performance} reported correlation coefficients of 0.11 and 0.06 in personalised and generalised protocols, respectively, using openSMILE for extracting audio features and OpenFace \citep{amos2016openface} for extracting video features, followed by a \ac{BiLSTM} network.

In addition to verbal and non-verbal features from audio, image and text, \citet{azari2019towards} experimented with a different type of feature: mutual or contagious laughter as a measure of synchrony between the speaker and listener during the interaction. \citet{hinduja2019fusion} used facial landmarks and spectrogram as hand-crafted features and \ac{CNN} output as deep features in a \ac{RF} model. Lastly, \citet{lusquino2020weightless} leveraged a different type of model -- \ac{WANN} -- and reported a correlation coefficient of 0.25 on the validation set of the \texttt{OMG-Empathy} dataset.

Other works in empathy detection as regression tasks primarily utilised classical \ac{ML} models (Appendix Table~B4). \citet{kroes2022empathizing} leveraged a \ac{LR} model on the \texttt{Human-VirtualAgent} dataset. With the \texttt{Teacher-Student} dataset, \citet{pan2022multimodal} comprehensively experimented with a wide range of features from audio and video in an AdaBoost model. Their extracted features include mid-level behavioural features -- such as facial expression, head pose and eye gaze -- and high-level interpretable features, such as video length, frequency of speaker switch and total number of words. Such feature extraction often yields good results but may require substantial computational resources and careful tuning to optimise the model.

\subsubsection{Classification Task (Level of Empathy)}
In classifying empathy levels, no studies used any common benchmark dataset (Appendix Table~B4). On the \texttt{DAIC-WOZ} dataset, \citet{tavabi2019multimodal} leveraged pre-trained \ac{BERT} to calculate text embedding and pre-trained \ac{ResNet} to calculate visual features in addition to action units and head pose features from OpenFace. As audio features, they extracted the extended Geneva minimalistic acoustic parameter set and \ac{MFCC} \citep{hasan2021many} using OpenSMILE. With these features, they experimented with \ac{GRU} and \ac{MLP} in different fusion techniques, where \ac{GRU}-based fusion of temporal audio and video sequences appeared to be the best fusion strategy in their setting, resulting an F1 score of 71\%. Their ablation experiment shows that text modality is more effective than video and audio modalities: text alone resulted in an F1 score of 64\%, whereas the video and audio individually provided F1 scores of 46\% and 38\%, respectively.

On the \texttt{MEDIC} dataset \citep{zhu2023medic}, the best result is achieved using SWAFN, a multimodal \ac{LSTM} network proposed by \citet{chen2020swafn}. The network uses three individual \acp{LSTM} to encode video, audio and textual modalities, followed by a novel aggregation strategy using a multi-task learning framework. Among the three mechanisms of the \texttt{MEDIC} dataset, the client's expression of experience was better classified than the counsellor's empathy \citep{zhu2023medic}, which supports the difficult nature of empathy detection compared to expression (i.e., emotion) recognition.

\subsection{Audio Signals}\label{sec:emp-aud}
Audio-based empathy detection works include audio from conversations in various contexts, such as healthcare and call centres. By audio-based empathy detection studies, we refer to studies that exclusively leverage audio and sometimes transcripts, which differ from multimodal audiovisual studies presented earlier.

Processing audio includes two primary approaches: directly utilising audio as a signal or converting it into text and employing text-based methods. As shown in Appendix Table~B5, none of the audio datasets are used across multiple studies to allow comparative analysis on model performance.

Given that audio-based datasets involve conversations between two persons, the works of \citet{chen2020automated} and \citet{xiao2015rate} include voice activity detection (`speech' or `no speech') and speaker diarisation (i.e., speaker separation) in their empathy detection workflow. \citet{chen2020automated,alam2016can,xiao2015rate} converted the audio into text sequences, followed by extracting features from the text sequences. \citet{chen2020automated} and \citet{alam2016can} extracted several lexical features, such as text embedding, from the audio transcripts and several acoustic features, such as \ac{MFCC}, from the audio signal. \citet{chen2020automated} reported better performance of lexical features than acoustic features. Lastly, \citet{xiao2015rate}'s empathy detection model on audio-based \texttt{CTT} dataset is entirely text-based -- leveraging uni-gram, bi-gram and tri-gram language models -- without audio-based features.

On the recent \texttt{MultimodalMI} dataset, \citet{tavabi2023therapist} used a distilled \ac{RoBERTa} pre-trained language model, a bidirectional \ac{GRU} layer followed by a two-head self-attention layer to predict a continuous empathy score between 0 and 1 (regression task). Using the same dataset, \citet{tran2023multimodal} proposes a multimodal empathy classification system utilising both audio and text transcripts to predict high vs low empathy. Features from the audio and texts are extracted using \ac{HuBERT} \citep{hsu2021hubert} and distilled \ac{RoBERTa} pre-trained models, respectively. The features are then passed through a bidirectional \ac{GRU} model, followed by modality fusion. They experimented with early and late fusion through \ac{MLP} layers. A wide range of experiments supports the effectiveness of late fusion in most experimental conditions, early fusion in some cases and text-only prediction in very few cases.

\subsection{Physiological Signals}
Research in physiological signal-based empathy detection typically adopts feature extraction, followed by \ac{ML} algorithms. Appendix Table~B6 reports the studies and methods of physiological signal-based empathy detection. No datasets are used across multiple studies.

With the \texttt{PainEmp} dataset, \citet{golbabaei2022physiological} extracted ten features and leveraged an \ac{SVM} with radial basis function kernel to detect cognitive and affective empathy. Lastly, \citet{kuijt2020prediction} extracted 15 features from the \texttt{\acs{EEG}} data and leveraged multiple \ac{LR} in the regression task and \ac{LR}, \ac{SVM} and \ac{DT} in the classification task. In the classification task, they only used five best-performing features. In both regression and classification settings, the participants' empathy before the experiment is better detected than `after' and `during' the experiment.

\subsection{Discussion: Findings, Challenges and Research Gaps}
\subsubsection{Lack of Benchmarking}
Benchmarking and comparative analysis of empathy detection models are hindered by a lack of common benchmark dataset usage, particularly in the domains of audio and physiological signals. Code availability further impacts benchmarking and the broader adoption of methodologies, as minute details of proposed algorithms are often not fully captured in the papers. Among 77 text-based empathy detection models (treating each dataset-specific implementation as a separate model), only 29 have shared their code. Similarly, for audiovisual, audio signal, and physiological signal-based models, the proportion of publicly accessible code is even lower, with only 7 out of 17, 2 out of 7, and none out of 4 releasing their implementations. Mandating the publication of code alongside research findings can ensure transparency and facilitate progress in the field.

A variety of evaluation metrics have been employed in the literature for empathy detection. For regression tasks predicting a continuous degree of empathy, metrics such as Pearson’s correlation, Spearman’s correlation, concordance, mean squared error, and R\textsuperscript{2} are commonly used. In classification tasks predicting discrete empathy levels, metrics include accuracy, F1 score, \ac{AUC}, average precision, unweighted average recall, and the Matthews correlation coefficient. However, inconsistent use of these metrics across studies can complicate cross-study comparisons. For instance, despite both using the \texttt{iEmpathize} dataset, the metrics reported in \citep{hosseini2021it} and \citep{hosseini2023feature} differ. While it is unreasonable to expect universal adoption of a single metric, efforts toward greater standardisation in evaluation frameworks and metric reporting would enhance comparability.

\subsubsection{Multimodal Empathy Detection}
The growth of empathy detection modalities, as illustrated in \autoref{fig:paper-vs-year}, shows a dominant rising trend in text-based empathy detection since 2020. However, the current body of research lacks equivalent development in audiovisual, audio, and physiological signals. Empathy detection systems based on these modalities can be particularly effective in scenarios where such signals are available and potentially provide a more comprehensive measure of empathy. While spoken information from video and audio can be converted to text for text-based empathy detection, video and audio contain additional information such as facial expressions and pitch. These elements can significantly enhance the accuracy and quality of empathy detection, which necessitates dedicated research in these areas.

A multimodal empathy detection system can effectively integrate these different types of data. Additionally, analysing the contributions of different modalities provides insights into the most important factors for an effective empathy detection system. Few studies \citep{tavabi2019multimodal,zhu2023medic,pan2022multimodal,tran2023multimodal,chen2024detecting} have shown proof of concept towards multimodal empathy detection. Overall, the multimodal approach holds promise for creating a robust empathy detection system by leveraging the strengths of various input modalities.

\subsubsection{LLM in Empathy Detection}
The recent success of \acp{LLM} presents an opportunity to utilise them in empathy detection tasks. \acp{LLM} can serve as the primary prediction model or as a supportive tool to enhance predictions made by conventional models. While \acp{LLM} may excel in empathy detection due to their extensive language understanding capabilities, their training and deployment often require substantial resources, which may be impractical for low-resource settings. Smaller optimised models like \ac{BERT} and \ac{RoBERTa} can offer reasonable performance with better resource efficiency and may be better suited for certain applications, such as remote areas with limited healthcare access, community counselling centres, education settings in low-income schools, and humanitarian aid and crisis response.

Even when not utilised as the primary prediction model, \acp{LLM} can contribute to empathy prediction tasks, particularly in data preprocessing tasks such as text rephrasing \citep{lu2023hit} and empathy annotation \citep{hasan2024llm-gem}. A recent study \citep{street2024llms} has shown that \acp{LLM} achieve human-level performance in theory of mind tasks. Drawing on the close relationship between cognitive empathy and theory of mind \citep{blair2005responding}, this indicates that \acp{LLM} possess (or can mimic) empathic skills that could potentially assist in empathy detection.

Multimodal \acp{LLM} hold promise for empathy detection in real-life audiovisual interactions, as suggested by \citet{hasan2024thesis}. This approach capitalises on \acp{LLM}’ advanced language understanding and multimodal abilities to interpret the nuances of natural conversations across audio, visual and text modalities. The emergence of multimodal \acp{LLM} -- such as OpenAI's \ac{GPT}-4o (omni), \ac{GPT}-4V (vision) and Google's Gemini -- enables both zero-shot (i.e., no fine-tuning) and few-shot (i.e., fine-tuning with a few examples) applications in empathy detection. A recent study on \ac{GPT}-4V \citep{lain2023gpt-4v} assessed its performance on 21 emotion recognition datasets across six tasks, including sentiment analysis, facial emotion recognition and multimodal emotion recognition, all in zero-shot settings. Although \ac{GPT}-4V demonstrated strong visual processing capabilities, it struggled with micro-expression recognition. Since \ac{GPT}-4V is designed primarily for general domains, future studies could focus on fine-tuning it in few-shot settings. Nonetheless, this proof of concept for multimodal \acp{LLM} in general emotion recognition indicates potential for empathy detection. With the rapid advancement of multimodal \acp{LLM}, they are likely to become a stronger candidate for empathy detection in the days to come.

\section{Applications of Empathy Detection}\label{sec:appl}
Empathy detection has the potential to bring transformative changes across various domains, such as healthcare, education and social media, Focusing on a few key domains, this section discusses potential benefits, associated challenges and ethical considerations.

\subsubsection*{Assessment of Communication Quality}
Empathy detection can be used to evaluate the quality of interpersonal communication, which can be used as feedback to improve interactions. Consider \textit{healthcare} as an example. A study on patient-doctor interaction found that 85\% of 563 patients either changed or were considering changing their doctors due to a lack of effective communication related to empathy being one of the main reasons \citep{cousins1985patients,bellet1991the}. Empathic doctors would be better equipped to communicate medical information in a fashion that the patient will attend to \citep{jani2012role}. Therefore, the service quality of healthcare providers could be assessed in terms of empathy if we could detect empathy in the first place. Assessment of healthcare providers can be in various contexts -- such as counselling sessions \citep{wu2021towards,gibson2015predicting,gibson2016deep,xiao2015rate}, oncology encounters \citep{chen2020automated}, and other general patient-doctor interactions \citep{shi2021modeling,dey2022enriching} -- either through telehealth or in-person. Measurement of empathy can also facilitate effective empathy training programs for healthcare professionals. In the same vein, empathy detection can be applied in educational interactions in teaching, customer service in businesses, and even in human-robot interactions.

Empathy detection can also improve communication quality in long-distance communication, such as that between international students and their families or during online interviews for jobs or university admissions. Video conferencing applications like Zoom and Microsoft Teams could incorporate live empathy feedback, similar to live transcripts, which is common nowadays. This way, people can see how their words and expressions are perceived in real-time, which may help people adjust their communication to be more empathic and responsive.

\subsubsection*{Disease Diagnosis}
Empathy detection systems can help diagnose diseases and cognitive disorders where a lack of empathy is a symptom, such as autism, psychopathy and alexithymia \citep{lamm2016shared}. Several studies have shown proof of concepts in this regard, such as diagnosing social communication disorders (Down syndrome, intellectual disability) \citep{hervas2016learning} and autism spectrum disorders \citep{golbabaei2022physiological}.

\subsubsection*{Social Media Moderation}
People often seek mental support through social media platforms. Accordingly, several works have detected empathy in various social media interactions, such as Reddit \citep{sharma2020computational}, Twitter \citep{hosseini2021distilling}, and cancer survivors networks \citep{khanpour2017identifying,hosseini2021it,hosseini2023feature}. We can envision a peer support platform where non-empathic responses are filtered out through an empathy detection system. This way, social media platforms can foster empathic responses while discouraging non-empathic ones.

\subsubsection*{Ethical Considerations and Challenges}
Visibility of empathy scores might encourage individuals to feign empathy, similar to how fake facial expressions are a concern in emotion detection \citep{hossain2020using}. Developing robust methods to differentiate genuine empathy from feigned responses will be crucial to ensure the reliability and effectiveness of empathy detection systems. Continuous feedback on empathic behaviour may also inadvertently create undue pressure on individuals, which may affect their mental health or authenticity in interactions. Safeguards should be in place to mitigate such unintended consequences.

What is considered a normal conversation and what is perceived as harsh can vary greatly across cultures. Ensuring fairness, therefore, becomes a significant challenge. Systems may inherit biases from training data that predominantly represent a specific demographic, resulting in unfair or inaccurate assessments towards other demographic or cultural groups. Addressing these biases is essential to ensure fairness and inclusivity. Before deploying an empathy detection system, it is imperative to rigorously evaluate the generalisability of the system across diverse populations. Apart from these, other ethical considerations and challenges, such as data privacy and consent, commonly associated with general Affective Computing tasks, must also be carefully addressed to ensure the responsible and ethical deployment of empathy detection systems.

\section{Conclusion}\label{sec:concl}
Empathy, an involuntary and vicarious reaction to emotional signals from another individual or their circumstances \citep{hoffman1978toward}, has emerged as a promising research area across several disciplines. Empathy detection in Computer Science, particularly through \ac{ML} methodologies, has grown substantially in recent years. This paper presents a rigorous, systematic literature review following relevant \ac{PRISMA} guidelines for reproducibility. Starting with an extensive search across ten scholarly databases, we select 82 papers after a thorough screening process, including abstract and full-text screening based on exclusion criteria. We discuss and group similar papers based on task formulations and \ac{ML} methodologies. We present a task formulation hierarchy with representative datasets and their details, such as data collection, experiment details, statistics, annotation protocol and public availability. To describe \ac{ML} methodologies, we group our findings based on four input modalities: text sequences, audiovisual data, audio signals and physiological signals. In each modality, we enumerate the algorithms used, their performance and code availability.

This review uncovers several new insights into the computational empathy domain and identifies critical avenues for future research and development. Exploring novel task formulations such as parallel and bidirectional empathy, particularly in group settings and global-level measurements, can advance our understanding of empathy in complex social interactions. Further research comparing self-annotation and third-party annotation under controlled conditions is necessary to determine the most appropriate annotation scheme for empathy detection. The limited public availability of datasets and codes poses significant challenges for reproducibility and benchmarking in the field. At the same time, the diversity and lack of standardisation in evaluation metrics complicate consistent model comparison. Physiological signals offer promising avenues for more accurate empathy detection. Finally, the potential of \acp{LLM} and multimodal approaches to enhance empathy detection systems presents exciting opportunities for future research. 

Studies we review in this paper operationalise the term `empathy' in varying ways. For example, some define empathy purely in cognitive dimension, such as understanding another’s perspective or capacity to comprehend \citep{buechel2018modeling}, and some further integrate behavioural or prosocial dimensions, such as offering emotional support \citep{dey2022enriching}. Future reviews may adopt a more critical lens to distinguish studies according to how they operationalise empathy.

\section*{Acknowledgement}
We thank A/Professor Susannah Soon at Curtin University for her comments on the initial version of this paper.

\section*{List of Acronyms}\label{sec:acro}
\begin{acronym}
    \acro{AUC}{Area Under the receiver operating characteristics Curve}
    \acro{BERT}{Bidirectional Encoder Representations from Transformers}
    \acro{BiLSTM}{Bidirectional \acs{LSTM}}
    \acro{CNN}{Convolutional Neural Network}
    \acro{DeBERTa}{Decoding-Enhanced \acs{BERT} with Disentangled Attention}
    \acro{DistilBERT}{Distilled \acs{BERT}}
    \acro{DL}{Deep Learning}
    \acro{DT}{Decision Tree}
    \acro{EBM}{Explainable Boosting Machine}
    \acro{EC}{Exclusion Criteria}
    \acro{ECG}{Electrocardiogram}
    \acro{EEG}{Electroencephalogram}
    \acro{fMRI}{functional Magnetic Resonance Imaging}
    \acro{GPT}{Generative Pre-trained Transformer}
    \acro{GRU}{Gated Recurrent Unit}
    \acro{HuBERT}{Hidden-Unit BERT}
    \acro{IRI}{Interpersonal Reactivity Index}
    \acro{LLM}{Large Language Model}
    \acro{LR}{Linear Regression}
    \acro{LogR}{Logistic Regression}
    \acro{LSTM}{Long Short-Term Memory}
    \acro{MFCC}{Mel-Frequency Cepstral Coefficients}
    \acro{MISC}{Motivational Interviewing Skill Code}
    \acro{MITI}{Motivational Interviewing Treatment Integrity}
    \acro{ML}{Machine Learning}
    \acro{MLP}{Multi Layer Perceptron}
    \acro{NB}{Na\"ive Bayes}
    \acro{NLP}{Natural Language Processing}
    \acro{PBC4cip}{Pattern-Based Classifier for Class Imbalance Problems}
    \acro{PRISMA}{Preferred Reporting Items for Systematic reviews and Meta-Analyses}
    \acro{ResNet}{Residual Network}
    \acro{RGCN}{Relational Graph Convolutional Network}
    \acro{RNN}{Recurrent Neural Network}
    \acro{RF}{Random Forest}
    \acro{RoBERTa}{Robustly Optimised \acs{BERT} Pretraining Approach}
    \acro{RR}{Ridge Regression}
    \acro{SVM}{Support Vector Machine}
    \acro{SBERT}{Sentence \acs{BERT}}
    \acro{WANN}{Weightless Artificial Neural Network}
    \acro{WASSA}{Workshop on Computational Approaches to Subjectivity, Sentiment \& Social Media Analysis}
\end{acronym}


\ifCLASSOPTIONcaptionsoff
  \newpage
\fi

\appendices


\begin{table*}[!t]
\caption{Initial Search Results with Details in All 10 Databases.}
\label{tab:search_res}
\centering
\footnotesize
\begin{tabularx}{\textwidth}{@{}
>{\centering\arraybackslash}p{0.1cm}
>{\raggedright\arraybackslash}p{4.7cm}
>{\raggedright\arraybackslash}p{7.6cm}
>{\centering\arraybackslash}X
>{\centering\arraybackslash}p{1.8cm}
>{\centering\arraybackslash}p{1.3cm}
@{}} \toprule
\textbf{SL} & \textbf{Database} & \textbf{Search condition} & \textbf{Items} & \textbf{Auto-filtering criteria} & \textbf{Post-filter items} \\ \midrule
1 & Scopus & Searched in title, abstract and keywords & 233 & EC4, EC5 & 198 \\
2 & Web of Science & Searched in title, abstract, keywords on all databases & 227 & EC4, EC5 & 183 \\
3 & ScienceDirect & Search engine did not support wildcard & 27 & EC4, EC5 & 16 \\
4 & IEEE Xplore & Searched in all metadata & 93 & EC4 & 84 \\
5 & ACM Guide to Computing Literature & Searched in abstracts & 25 & EC4 & 22 \\
6 & dblp & Combined dblp search; search string: empath (detect $\vert$ recog) & 37 & -- & 37 \\
7 & Google Scholar & Sorted by relevance & 18,100 & EC4, First 100 & 100 \\
8 & PubMed & Searched in all fields & 55 & EC4 & 51 \\
9 & ProQuest & Searched in abstracts & 93 & EC4 & 88 \\
10 & ACL Anthology & Searched in title and abstract & 22 & -- & 22 \\
\bottomrule
\end{tabularx}
\end{table*}

\section{Paper Selection}\label{sec:paper-selec}
The systematic nature of this review ensures reproducibility and transparency in the selection and analysis of relevant papers. We adhered to the relevant recommendations from the PRISMA 2020 guidelines \citep{page2021prisma} as well as published systematic reviews in Affective Computing \citep{pampouchidou2019automatic-slr,aranha2021adapting-slr,ma2023a-slr,pepa2023automatic-slr,saganowski2023emotion-slr,jacobs2024threat-slr}. For example, our paper selection process considered the following inclusion and exclusion criteria.

\subsection*{Inclusion Criteria (IC)}
\begin{enumerate}[label={IC\arabic*.}, left=0pt]
    \item Detect empathy using any \ac{ML} algorithms
    \item Peer-reviewed full-length research paper
    \item Published between January 2013 and June 2024
\end{enumerate}

\subsection*{Exclusion Criteria (EC)}
\begin{enumerate}[label={EC\arabic*.}, left=0pt]
    \item Not a full-length research paper (e.g., conference abstracts and conference proceeding books)
    \item No use of artificial intelligence, machine learning or deep learning
    \item Review, survey, meta-analysis, thesis or dissertation
    \item Not in English
\end{enumerate}

We included papers from 2013 onward to capture the period when \ac{ML}, and particularly \ac{DL}, became widely feasible. Key developments, such as the first modern CNN (AlexNet) in 2012 \citep{krizhevsky2012imagenet} and optimisation algorithms like Adam \citep{kingma2015adam} in 2014-2015, led \ac{DL}’s success on a variety of tasks \citep{zhang2023dive,lecun2015deep}. Our systematic search validated this time frame, as we found no relevant empathy detection studies published in 2013 or 2014.

\subsection{Paper Search}
We formulate a search string using logical operators (AND and OR) among synonymous terms of empathy, detection and artificial intelligence: empath* AND (detect* OR recog*) AND (``deep learning'' OR ``machine learning'' OR ``artificial intelligence'' OR AI). The asterisk (*) is a wildcard character that facilitates the inclusion of any number of characters in place of the asterisk.

With the search string, one researcher (MRH) searched for relevant records across ten databases (see \autoref{tab:search_res} for more details) on 24 February 2023. Among the search engines, ACL Anthology does not support logical search. We, therefore, built a program\footnote{\url{https://github.com/hasan-rakibul/boolean-search-bib-abstract}} to search in the ACL database using the available bibliography document. Several search engines, such as Scopus and Web of Science, support filtering based on publication year (\acs{IC}3) and paper type (\acs{EC}1 and \acs{EC}3), so we automatically filtered out the search results. \autoref{tab:search_res} presents the number of search results, search condition (e.g., title, abstract, full-paper, etc.), automatic-filtering results and corresponding filtering criteria.

\subsection{Paper Screening}
\begin{figure}[!t]
    \centering
    \includegraphics[width=0.95\columnwidth]{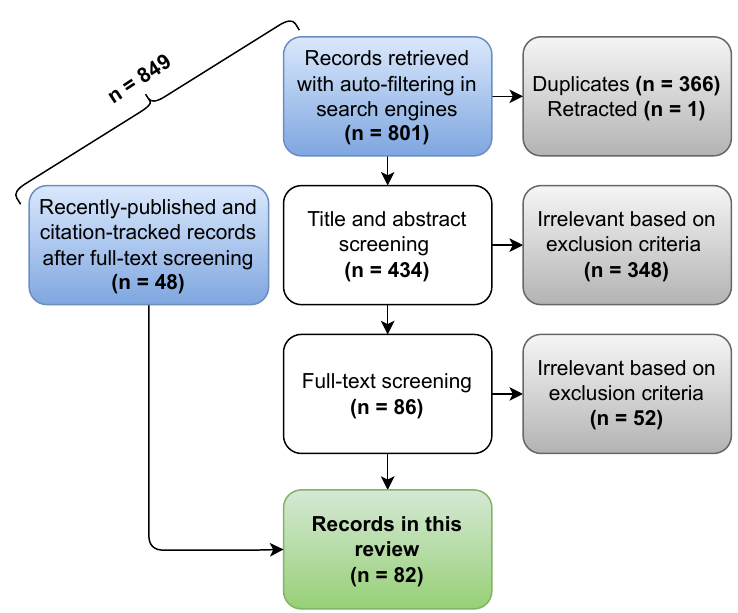}
    \caption{Number of records at different stages in our screening process.}
    \label{fig:prisma}
\end{figure}

\autoref{fig:prisma} illustrates our step-by-step paper screening strategy. We initially gathered 801 records from the databases. After removing duplicated and retracted records, one researcher (MRH) screened the remaining 434 records by reading titles and abstracts using the Covidence systematic review management software \citep{covidence}. At this stage, records were excluded only if they clearly met one or more \ac{EC}. Records that could not be conclusively evaluated based on the title and abstract proceeded to the full-text screening stage. At this stage, three researchers (MRH, MZH and SG) screened the 86 remaining records. Some records were deemed ambiguous for inclusion or exclusion, and these were discussed among all researchers in regular weekly meetings during the full-text screening period until a consensus was reached. For instance, while \citet{hossain2022detection} and \citet{hinduja2019mitigating} initially appeared to meet the inclusion criteria, further examination revealed that \citet{hossain2022detection} perform sentiment analysis on how people react toward online reviews, and \citet{hinduja2019mitigating} analyses potential biases in empathy detection without detecting any types of empathy.

\begin{table}[!t]
    \centering
    \scriptsize
    \begin{threeparttable}
    \caption{Summary of Empathy Detection Studies Modelled as a \textit{Regression} Task (Degree of Empathy) on Text Datasets Used in a Single Published Work.}
    \label{tab:study_text_regr_app}
    \begin{tabularx}{\columnwidth}{@{}
    >{\raggedright\arraybackslash}p{2.0cm}
    >{\centering\arraybackslash}p{0.6cm}
    >{\raggedright\arraybackslash}p{1.3cm}
    >{\raggedright\arraybackslash}p{2.5cm}
    >{\centering\arraybackslash}X
    @{}} \toprule
    \textbf{Dataset} & \textbf{Study} & \textbf{Best Model} & \textbf{Performance\tnote{a}} & \textbf{Code Avail.\tnote{a}} \\ \midrule
    MI & \citep{gibson2015predicting} & \acs{LR} & SCC: 0.611 & $\times$ \\
    PathogenicEmp & \citep{abdul2017recognizing} & \acs{RR} & PCC: 0.252 & $\times$ \\
    LeadEmpathy & \citep{sedefoglu2024leadempathy} & \acs{BERT} & PCC: 0.816 & \checkmark \\
    EmpathicStories & \citep{shen-etal-2023-modeling} & \acs{SBERT} & PCC: 0.309, SCC: 0.352 & \checkmark \\
    NewsEssay v4+v3 & \citep{soni2025evaluation} & \acs{RoBERTa} & PCC: 0.77 & \checkmark \\
    \bottomrule
    \end{tabularx}
    \begin{tablenotes}
        \item[a] PCC -- Pearson correlation coefficient; SCC -- Spearman’s correlation coefficient
    \end{tablenotes}
    \end{threeparttable}
\end{table}

\begin{table*}[!thb]
    \centering
    \footnotesize
    \begin{threeparttable}
    \caption{Summary of Empathy Detection Studies Modelled as a \textit{Classification} Task (Level of Empathy) on Text Datasets Used in a Single Published Work.}
    \label{tab:study_text_class_app}
    \begin{tabularx}{\textwidth}{
    >{\raggedright\arraybackslash}p{4.9cm}
    >{\centering\arraybackslash}p{0.6cm}
    >{\raggedright\arraybackslash}p{3.5cm}
    >{\raggedright\arraybackslash}p{6.5cm}
    >{\centering\arraybackslash}X
    } \toprule
    \textbf{Dataset} & \textbf{Study} & \textbf{Best Model} & \textbf{Performance\tnote{a}} & \textbf{Code Avail.} \\ \midrule
    MedicalCare + NewsEssay & \citep{shi2021modeling} & \acs{SVM} & Acc: $89.4\%$, F1: $78.4\%$ & $\times$\\
    MedicalCare v2 & \citep{dey2022enriching} & \acs{BERT} & F1 $\in [75\%, 85\%]$ & $\times$\\
    MedicalCare v3 & \citep{dey2023investigating} & \acs{BERT} & F1 $\in [66\%, 75\%]$ & $\times$\\
    AcnEmpathize & \citep{lee2024acnempathize} & \acs{DistilBERT}, \acs{RoBERTa} & Acc $89.3\%$, F1 $ \in [77.4\%, 93.1\%]$ \\
    LeadEmpathy & \citep{sedefoglu2024leadempathy} & \acs{SVM} (Binary), \acs{BERT} (Multi) & F1: $81.7\%$ (Binary), $49.9\%$ (Multi) & $\times$\\
    MI & \citep{gibson2015predicting} & \acs{NB} & Unweighted average recall: $75.3\%$ & $\times$\\
    MI v2 & \citep{gibson2016deep} & \acs{MLP}-\acs{LSTM} & Unweighted average recall: $79.6\%$ & $\times$\\
    LungBreastCSN & \citep{khanpour2017identifying} & \acs{CNN}-\acs{LSTM} & F1: $78.4\%$ & $\times$\\
    PEC, EmpathicDialogues, RolePlayMI & \citep{wu2021towards} & \acs{BERT} & Matthews correlation coefficient $\in [\approx 0.56, \approx 0.95]$ & $\times$\\
    EmpathicDialogues v2 & \citep{montiel2022explainable} & \acs{PBC4cip} & \acs{AUC}: $62.5\%$ & $\times$\\
    NewsEssay & \citep{hosseini2021distilling} & \acs{BERT}-\acs{MLP} & F1: $68.4\%$ & $\times$\\
    EPITOME v2, NewsEssay & \citep{hosseini2022calibrating} & \acs{BERT}, \acs{RoBERTa} & Acc $\in [61.5\%, 71.8\%]$ & $\times$\\
    Brand-Customer & \citep{singh2022linguistic} & \acs{RoBERTa} & F1: $73\%$ & $\times$\\
    FacebookReview & \citep{arahim2021assessing} & \acs{SVM} & Acc: $21.5\%$, F1: $75.7\%$ & $\times$\\
    NewsConvD & \citep{lee-etal-2024-empatheticfig} & mT5-\acs{BERT}-\acs{MLP} & PCC: 0.012 & $\times$ \\
    \bottomrule
    \end{tabularx}
    \begin{tablenotes}
        \item[a] Range of performance is reported when overall classification performance is unavailable
    \end{tablenotes}
    \end{threeparttable}
\end{table*}

\begin{table*}[!t]
    \centering
    \footnotesize
    \begin{threeparttable}
    \caption{Summary of Empathy Detection Studies on Audiovisual Datasets Used in a Single Published Work.}
    \label{tab:study_video_app}
    \begin{tabularx}{\textwidth}{
    >{\raggedright\arraybackslash}p{2.8cm}
    >{\centering\arraybackslash}p{1.5cm}
    >{\raggedright\arraybackslash}p{4.5cm}
    >{\raggedright\arraybackslash}p{5.5cm}
    >{\centering\arraybackslash}X
    } \toprule
    \textbf{Dataset} & \textbf{Study} & \textbf{Best Model} & \textbf{Performance}\tnote{a} & \textbf{Code Avail.\tnote{b}} \\ \midrule
    
\multicolumn{5}{@{}l}{\texttt{\textbf{Regression}}} \\
    Human-VirtualAgent & \citep{kroes2022empathizing} & \acs{LR} & R\textsuperscript{2}: $0.485$ & $\times$ \\ 
    Teacher-Student & \citep{pan2022multimodal} & AdaBoost & MSE: $0.374$ & \checkmark \\
    EmpathicStories++ & \citep{shen-etal-2024-empathicstories} & \acs{GPT}4 & PCC: 0.232, SCC: 0.176 & $\times$ \\ 
    MultimodalMI & \citep{tavabi2023therapist} & \acs{RoBERTa}-\acs{GRU} & CCC $\in [0.408, 0.596]$ & \checkmark \\ \midrule
\multicolumn{5}{@{}l}{\texttt{\textbf{Classification}}} \\
    Teacher-Student & \citep{pan2022multimodal} & \acs{DT} & Acc: $90.9\%$, F1: $90.1\%$ & \checkmark \\ 
    Human-Robot & \citep{mathur2021modeling} & XGBoost & Acc: $69\%$, \acs{AUC}: $72\%$ & $\times$ \\
    Human-Avatar & \citep{hervas2016learning} & \acs{LogR} & F1 $\in [72\%, 78\%] $ & $\times$ \\
    DAIC-WOZ & \citep{tavabi2019multimodal} & \acs{ResNet}, \acs{BERT}, \acs{GRU}, \acs{MLP} & F1: $71\%$ & $\times$ \\
    MEDIC & \citep{zhu2023medic} & \acs{LSTM} & Acc $\in [77.6\%, 86.4\%]$, F1 $\in [77.7\%, 86.3\%]$ & $\times$ \\
    \textit{Various online sources}\tnote{b} & \citep{alanazi2023prediction} & \acs{CNN} & Acc: $98.9\%$, \acs{AUC}: $99\%$, F1: $91\%$ & $\times$ \\
    \bottomrule
    \end{tabularx}
    \begin{tablenotes}
        \item[a] Range of performance is reported when overall performance is unavailable. PCC -- Pearson correlation coefficient; CCC -- Concordance Correlation Coefficient
        \item[b] Description of the dataset, such as the number of samples and ground truth label space, is unavailable in the paper
    \end{tablenotes}
    \end{threeparttable}
\end{table*}

\begin{table}[!t]
    \centering
    \scriptsize
    \begin{threeparttable}
    \caption{Summary of Empathy Detection Studies on Audio Datasets.}
    \label{tab:study_audio}
    \begin{tabularx}{\columnwidth}{@{}
    >{\raggedright\arraybackslash}p{1.3cm}
    >{\centering\arraybackslash}p{0.6cm}@{$\quad$}
    >{\raggedright\arraybackslash}p{2.2cm}@{$\quad$}
    >{\raggedright\arraybackslash}p{3.2cm}@{}
    >{\centering\arraybackslash}X
    @{}} \toprule
    \textbf{Dataset} & \textbf{Study} & \textbf{Best Model} & \textbf{Performance\tnote{a}} & \textbf{Code Avail.} \\ \midrule    
\multicolumn{5}{@{}l}{\texttt{\textbf{Regression}}} \\
    CTT & \citep{xiao2015rate} & \acs{LR} & PCC $\in [0.65, 0.71]$ & $\times$ \\ \midrule
\multicolumn{5}{@{}l}{\texttt{\textbf{Classification}}} \\
    CTT & \citep{xiao2015rate} & \acs{SVM} & Acc $\in [80.5\%, 89.9\%]$, F1 $\in [85.3\%, 90.3\%]$ & $\times$ \\ \addlinespace
    COPE & \citep{chen2020automated} & \acs{SVM} & Avg. precision: $7.61\%$ & $\times$ \\
    CallCentre & \citep{alam2016can} & \acs{SVM} & Unweighted avg. recall: $65.1\%$ & $\times$ \\
    MultimodalMI & \citep{tran2023multimodal} & \acs{RoBERTa}, \acs{HuBERT}, \acs{GRU}, \acs{MLP} & F1 $\in [58.3\%, 72.6\%]$ & \checkmark \\ 
    SpeechEmpathy & \citep{chen2024detecting} & \acs{RoBERTa}, \acs{MLP} & Acc: 0.781, F1: 0.840 & \checkmark \\ 
    OTLA & \citep{sanjeewa2025machine} & \acs{LogR} & AUC: 0.617 & $\times$ \\
    \bottomrule
    \end{tabularx}
    \begin{tablenotes}
        \item[a] Range of performance is reported when overall performance is unavailable.
    \end{tablenotes}
    \end{threeparttable}
\end{table}

\begin{table}[!t]
    \centering
    \scriptsize
    \begin{threeparttable}
    \caption{Summary of Empathy Detection Studies on Physiological Datasets.}
    \label{tab:study_physio}
    \begin{tabularx}{\columnwidth}{@{}
    >{\raggedright\arraybackslash}p{0.6cm}
    >{\centering\arraybackslash}p{0.6cm}
    >{\raggedright\arraybackslash}p{1.1cm}
    >{\raggedright\arraybackslash}p{4.2cm}
    >{\centering\arraybackslash}X
    @{}} \toprule
    \textbf{Dataset} & \textbf{Study} & \textbf{Best Model} & \textbf{Performance\tnote{a}} & \textbf{Code Avail.} \\ \midrule
    
    \multicolumn{5}{@{}l}{\texttt{\textbf{Regression}}} \\
    fMRI & \citep{wei2021effective} & \acs{LR} & Pearson correlation: $0.54$, MSE: $20.1$ & $\times$ \\
    \ac{EEG} & \citep{kuijt2020prediction} & \acs{LR} & MSE $\in [51.749, 150.556]$ & $\times$ \\ \midrule

    \multicolumn{5}{@{}l}{\texttt{\textbf{Classification}}} \\
    PainEmp & \citep{golbabaei2022physiological} & \acs{SVM} & Acc $\in [79\%, 84\%]$ & $\times$ \\
    \ac{EEG} & \citep{kuijt2020prediction} & \acs{SVM}, \acs{DT} & Acc $\in [61.8\%, 74.2\%]$, F1 $\in [61.5\%, 74.3\%]$ & $\times$ \\
    \bottomrule
    \end{tabularx}
    \begin{tablenotes}
        \item[a] Range of performance is reported when overall performance is unavailable.
    \end{tablenotes}
    \end{threeparttable}
\end{table}

\begin{sidewaystable*}
\centering
\scriptsize
\begin{threeparttable}
\caption{PRISMA 2020 Checklist.}\label{tab:prisma}
\begin{tabularx}{\linewidth}{
>{\raggedright\arraybackslash}p{2.8cm}
>{\centering\arraybackslash}p{0.5cm}
>{\raggedright\arraybackslash}X
>{\raggedright\arraybackslash}p{2.2cm}}
\cellcolor[HTML]{AAAAE1}Section and Topic & \cellcolor[HTML]{AAAAE1}Item \# & \cellcolor[HTML]{AAAAE1}Checklist item & \cellcolor[HTML]{AAAAE1}Location where item is reported \\
\multicolumn{3}{l}{\cellcolor[HTML]{FFFFCC}\textbf{TITLE}} & \cellcolor[HTML]{FFFFCC} \\
Title & 1 & Identify the report as a systematic review. & Title \\
\multicolumn{3}{l}{\cellcolor[HTML]{FFFFCC}\textbf{ABSTRACT}} & \cellcolor[HTML]{FFFFCC} \\
Abstract & 2 & See the PRISMA 2020 for Abstracts checklist. & Abstract \\
\multicolumn{3}{l}{\cellcolor[HTML]{FFFFCC}\textbf{INTRODUCTION}} & \cellcolor[HTML]{FFFFCC} \\
Rationale & 3 & Describe the rationale for the review in the context of existing knowledge. & Section 1 \\
Objectives & 4 & Provide an explicit statement of the objective(s) or question(s) the review addresses. & Section 1 \\
\multicolumn{3}{l}{\cellcolor[HTML]{FFFFCC}\textbf{METHODS}} & \cellcolor[HTML]{FFFFCC} \\
Eligibility criteria & 5 & Specify the inclusion and exclusion criteria for the review and how studies were grouped for the syntheses. & Appendices \& Section 3 \\
Information sources & 6 & Specify all databases, registers, websites, organisations, reference lists and other sources searched or consulted to identify studies. Specify the date when each source was last searched or consulted. & Appendices \\
Search strategy & 7 & Present the full search strategies for all databases, registers and websites, including any filters and limits used. & Appendices \\
Selection process & 8 & Specify the methods used to decide whether a study met the inclusion criteria of the review, including how many reviewers screened each record and each report retrieved, whether they worked independently, and if applicable, details of automation tools used in the process. & Appendices \\
Data collection process & 9 & Specify the methods used to collect data from reports, including how many reviewers collected data from each report, whether they worked independently, any processes for obtaining or confirming data from study investigators, and if applicable, details of automation tools used in the process. & Appendices \\
 & 10a & List and define all outcomes for which data were sought. Specify whether all results that were compatible with each outcome domain in each study were sought (e.g. for all measures, time points, analyses), and if not, the methods used to decide which results to collect. & Sections 3 \& 4 \\
\multirow{-2}{*}{Data items} & 10b & List and define all other variables for which data were sought (e.g. participant and intervention characteristics, funding sources). Describe any assumptions made about any missing or unclear information. & Sections 3 \\
Study risk of bias assessment & 11 & Specify the methods used to assess risk of bias in the included studies, including details of the tool(s) used, how many reviewers assessed each study and whether they worked independently, and if applicable, details of automation tools used in the process. & Appendices \\
Effect measures & 12 & Specify for each outcome the effect measure(s) (e.g. risk ratio, mean difference) used in the synthesis or presentation of results. & Sections 3 \& 4 \\
 & 13a & Describe the processes used to decide which studies were eligible for each synthesis (e.g. tabulating the study intervention characteristics and comparing against the planned groups for each synthesis (item \#5)). & Appendices \\
 & 13b & Describe any methods required to prepare the data for presentation or synthesis, such as handling of missing summary statistics, or data conversions. & Sections 3 \& 4 \\
 & 13c & Describe any methods used to tabulate or visually display results of individual studies and syntheses. & Sections 3 \& 4 \\
 & 13d & Describe any methods used to synthesize results and provide a rationale for the choice(s). If meta-analysis was performed, describe the model(s), method(s) to identify the presence and extent of statistical heterogeneity, and software package(s) used. & Sections 3 \& 4 \\
 & 13e & Describe any methods used to explore possible causes of heterogeneity among study results (e.g. subgroup analysis, meta-regression). & Sections 3 \& 4 \\
\multirow{-6}{*}{Synthesis methods} & 13f & Describe any sensitivity analyses conducted to assess robustness of the synthesized results. & Not applicable \\
Reporting bias assessment & 14 & Describe any methods used to assess risk of bias due to missing results in a synthesis (arising from reporting biases). & Appendices \\
Certainty assessment & 15 & Describe any methods used to assess certainty (or confidence) in the body of evidence for an outcome. & Sections 3 \& 4 \\
\multicolumn{3}{l}{\cellcolor[HTML]{FFFFCC}\textbf{RESULTS}} & \cellcolor[HTML]{FFFFCC} \\
 & 16a & Describe the results of the search and selection process, from the number of records identified in the search to the number of studies included in the review, ideally using a flow diagram. & Appendices \\
\multirow{-2}{*}{Study selection} & 16b & Cite studies that might appear to meet the inclusion criteria, but which were excluded, and explain why they were excluded. & Appendices \\
Study characteristics & 17 & Cite each included study and present its characteristics. & Sections 3 \& 4 \\
Risk of bias in studies & 18 & Present assessments of risk of bias for each included study. & Not applicable \\
Results of individual studies & 19 & For all outcomes, present, for each study: (a) summary statistics for each group (where appropriate) and (b) an effect estimate and its precision (e.g. confidence/credible interval), ideally using structured tables or plots. & Sections 3 \& 4 \\
 & 20a & For each synthesis, briefly summarise the characteristics and risk of bias among contributing studies. & Sections 3 \& 4 \\
 & 20b & Present results of all statistical syntheses conducted. If meta-analysis was done, present for each the summary estimate and its precision (e.g. confidence/credible interval) and measures of statistical heterogeneity. If comparing groups, describe the direction of the effect. & Sections 3 \& 4 \\
 & 20c & Present results of all investigations of possible causes of heterogeneity among study results. & Sections 3 \& 4 \\
\multirow{-4}{*}{Results of syntheses} & 20d & Present results of all sensitivity analyses conducted to assess the robustness of the synthesized results. & Not applicable \\
Reporting biases & 21 & Present assessments of risk of bias due to missing results (arising from reporting biases) for each synthesis assessed. & Appendices \\
Certainty of evidence & 22 & Present assessments of certainty (or confidence) in the body of evidence for each outcome assessed. & Section 3 \\
\multicolumn{3}{l}{\cellcolor[HTML]{FFFFCC}\textbf{DISCUSSION}} & \cellcolor[HTML]{FFFFCC} \\
 & 23a & Provide a general interpretation of the results in the context of other evidence. & Sections 3, 4 \& 5 \\
 & 23b & Discuss any limitations of the evidence included in the review. & Sections 3 \& 4 \\
 & 23c & Discuss any limitations of the review processes used. & Appendices \\
\multirow{-4}{*}{Discussion} & 23d & Discuss implications of the results for practice, policy, and future research. & Sections 3, 4, 5 \& 6 \\
\multicolumn{3}{l}{\cellcolor[HTML]{FFFFCC}\textbf{OTHER INFORMATION}} & \cellcolor[HTML]{FFFFCC} \\
 & 24a & Provide registration information for the review, including register name and registration number, or state that the review was not registered. & Not applicable \\
 & 24b & Indicate where the review protocol can be accessed, or state that a protocol was not prepared. & Not applicable \\
\multirow{-3}{*}{Registration and protocol} & 24c & Describe and explain any amendments to information provided at registration or in the protocol. & Not applicable \\
Support & 25 & Describe sources of financial or non-financial support for the review, and the role of the funders or sponsors in the review. & Article Submission System \\
Competing interests & 26 & Declare any competing interests of review authors. & Article Submission System \\
Availability of data, code and other materials & 27 & Report which of the following are publicly available and where they can be found: template data collection forms; data extracted from included studies; data used for all analyses; analytic code; any other materials used in the review. & Appendices \\ \bottomrule
\end{tabularx}
\begin{tablenotes}
    \item \textit{From:} \citet{page2021prisma}. This work is licensed under CC BY 4.0. To view a copy of this license, visit \url{https://creativecommons.org/licenses/by/4.0/}
\end{tablenotes}
\end{threeparttable}
\end{sidewaystable*}

We screened another 27 recent papers, which we received through notifications and `snowballing'. Several search engines, such as Scopus, Web of Science, IEEE Xplore, ACM and Google Scholar, offer email notification services based on a predefined search string. Our `snowball' search involves examining reference lists of the included papers to identify potential new records. In an update on 06 June 2024, we incorporated eight newly published papers. In our latest update on 07 May 2025, we added 20 recent papers, which brought the total number of relevant papers in this systematic review to 82. MRH, in consultation with other researchers (co-authors) in this study, extracted relevant information from these selected papers. We categorise the analysis of the selected papers based on task formulations and data modality: text, audiovisual, audio and physiological signals.

Overall, the selection and analysis of papers were systematic and transparent, involving automated searches across multiple databases, clearly defined inclusion and exclusion criteria, and collaborative decision-making. This definitive and collaborative approach minimised the likelihood of individual bias. However, potential biases may still arise from limitations in the selected databases, which might not capture all relevant studies, as well as from the interpretation of ambiguous records by the researchers. 

\section{Modality-Specific Empathy Detection Methods}\label{sec:isolated}
This section lists studies based on datasets that have each been used in only one published work to date. 

\subsection{Text Sequence}
\subsubsection{Regression Task (Degree of Empathy)}
On continuous degrees of empathy detection on text datasets (\autoref{tab:study_text_regr_app}), works include therapists' empathy detection on \texttt{MI} dataset \citep{gibson2015predicting} and pathogenic empathy detection on social media \citep{abdul2017recognizing}. Both of them leveraged classical \ac{ML} methods: \ac{LR} and \ac{RR}. Classical \acp{ML} require fewer computational resources but often underperform transformer-based \ac{DL} algorithms, and as such, future research may explore recent algorithms, such as transformers, with these datasets.

\subsubsection{Classification Task (Level of Empathy)}

Studies presented in \autoref{tab:study_text_class_app} may not allow performance comparison between studies. However, a distinction can be made between classical \ac{ML} and \ac{DL}-based language model usage. For example, on detecting empathy in medical essays, \citet{shi2021modeling} experimented with \ac{SVM} and \ac{NB} on \texttt{MedicalCare} dataset, yielding an F1 score of 78.4\%. In \texttt{MedicalCare v2} dataset, \citet{dey2022enriching} experimented with \ac{BERT}, \ac{RoBERTa}, \ac{SVM}, \ac{NB}, \ac{LogR}, \ac{LSTM} and \ac{BiLSTM} models and reported an F1 score of 85\%. Similar experiments are also conducted with the \texttt{MedicalCare v3} dataset \citep{dey2023investigating}. Both experiments reveal the superior performance of \ac{BERT}-based models. Incorporation of FrameNet pre-trained model \citep{baker1998berkeley} boosted the baseline performance in the \texttt{v2} dataset \citep{dey2022enriching}. In the \texttt{v3} dataset \citep{dey2023investigating}, various linguistic constructions -- such as active or passive voice, static or energetic tone -- enhanced binary empathy classification performance compared to the baseline \ac{BERT} model.

\citet{lee2024acnempathize} experimented with two classical \ac{ML} algorithms (\ac{NB} and \ac{LogR}) and four pre-trained language model (including \ac{BERT}, \ac{RoBERTa} and \ac{DistilBERT}) on the \texttt{AcnEmpathize} dataset. Among these, \ac{DistilBERT} resulted in the best overall accuracy of 89.3\%, while the accuracy of \ac{BERT} and \ac{RoBERTa} was also close: 89.1\% and 88.5\%, respectively. Due to the unbalanced nature of the dataset, the authors also reported class-wise precision, recall and F1 scores. Although \ac{NB} provided better precision in the empathy class and better recall in the no-empathy class, it underperforms the \ac{BERT}-based models in the precision and F1 scores. The best F1 scores of 77.4\% in the empathy class and 93.1\% in the no empathy class are achieved by \ac{RoBERTa} and \ac{DistilBERT}, respectively.

On the \texttt{LeadEmpathy} dataset, \citet{sedefoglu2024leadempathy} employed \ac{SVM} for binary classification and \ac{BERT} for a 10-class classification task. The 10-class F1 score was notably lower at 45.7\%, reflecting the challenge of fine-grained empathy classification compared to the more straightforward binary classification, which achieved a higher F1 score of 81.7\%.

Instead of language models, several studies leveraged traditional \ac{DL} models like \ac{LSTM} and \ac{CNN}. \citet{gibson2015predicting} reported \ac{NB} as the optimal model in \texttt{MI} dataset. In a later study, \citet{gibson2016deep} reported that a combination of \ac{MLP} and \ac{LSTM} are the optimal model in the closely related \texttt{MI v2} dataset, yielding a higher unweighted average recall from 75.3\% to 79.6\%. \citet{khanpour2017identifying} used a combination of \ac{CNN} and \ac{LSTM} on the \texttt{LungBreastCSN} dataset. Other than commonly known models, \citet{montiel2022explainable} reported \ac{PBC4cip} -- exclusively designed for imbalanced datasets -- as the most effective classifier compared to several classical \ac{ML} baselines on \texttt{EmpathicDialogues v2} dataset.

\citet{hosseini2021distilling,hosseini2022calibrating} used knowledge distillation, which refers to the process of transferring knowledge from a large, complex model (teacher) to a smaller, simpler model (student) to improve the latter's performance while maintaining efficiency. \citet{hosseini2022calibrating} used \texttt{EPITOME v2} as an in-domain dataset and \texttt{NewsEssay} as an out-of-domain dataset to transfer knowledge from a \ac{RoBERTa} teacher model to a \ac{RoBERTa} student model. Their knowledge distillation framework boosted the performance compared to \ac{BERT} and \ac{RoBERTa} baselines. In their study, the \texttt{NewsEssay} dataset was used in a binary classification setting instead of the dataset's default usage as a regression task. Such a binary classification setup is also utilised by \citet{shi2021modeling} and \citet{hosseini2021distilling} using \ac{SVM} and \ac{BERT}-\ac{MLP} models, respectively. 

\subsection{Audiovisual Content}
As presented in \autoref{tab:study_video_app}, most studies leveraged a variety of classical \ac{ML} algorithms. For example, \citet{mathur2021modeling} experimented with eight classical \ac{ML} and two \ac{DL} models and reported XGBoost as the best model on the \texttt{Human-Robot} dataset.

\subsection{Audio Signals}
Most of the studies on audio datasets (\autoref{tab:study_audio}) reported classical \ac{ML} algorithms as the best in corresponding experiments: \ac{SVM} in empathy classification on the \texttt{CTT} \citep{xiao2015rate}, \texttt{COPE} \citep{chen2020automated} and \texttt{CallCentre} \citep{alam2016can} datasets and \ac{LR} in the regression study on the \texttt{CTT} dataset by \citet{xiao2015rate}.

\subsection{Physiological Signals}
All empathy detection studies on physiological signals (\autoref{tab:study_physio}) leveraged classical \ac{ML} algorithms: \ac{LR} and \ac{SVM}, each in two studies.

\section{PRISMA Checklist}\label{sec:app-prisma}
\autoref{fig:prisma} presents the completed PRISMA 2020 checklist \citep{page2021prisma}.

\printbibliography


%
\begin{IEEEbiography}[{\includegraphics[width=1in,height=1.25in,clip,keepaspectratio]{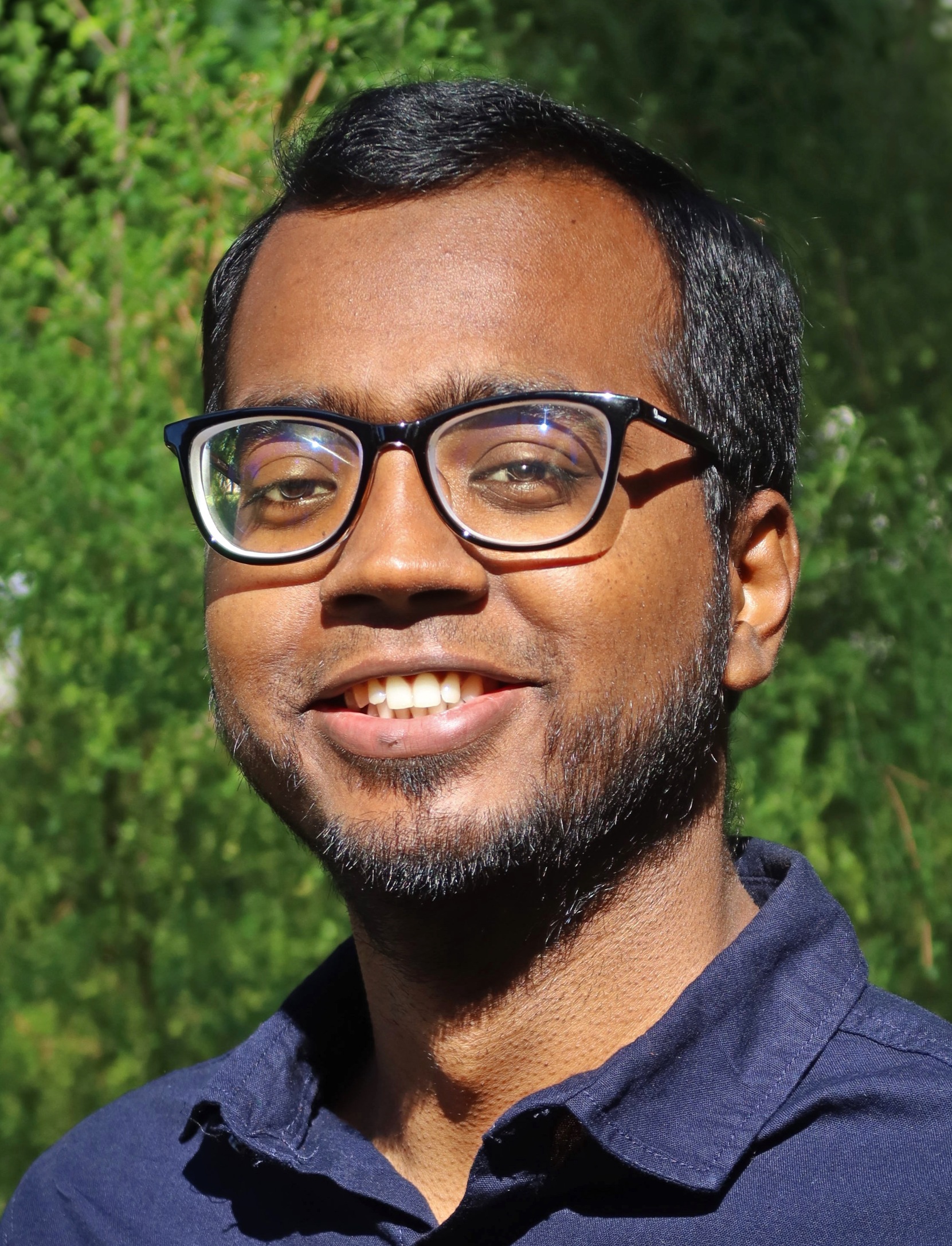}}]{Md Rakibul Hasan}
received his BSc (Hons) (2019) and MSc (2021) degrees from Khulna University of Engineering \& Technology, Bangladesh. Currently, he is a PhD candidate in Computing at Curtin University, Western Australia, where he builds deep learning models to detect empathy from multimodal data. His overarching research interest includes advancing deep learning algorithms for natural language processing and computer vision.

\end{IEEEbiography}
\begin{IEEEbiography}[{\includegraphics[width=1in,height=1.25in,clip,keepaspectratio]{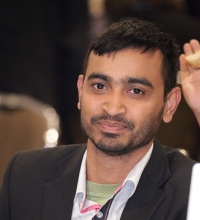}}]{Md Zakir Hossain}
completed the BSc (2011) and MSc (2014) from Khulna University of Engineering \& Technology (KUET), Bangladesh, and PhD (2019) from the Australian National University (ANU). He is a Senior Research Fellow at Curtin University. His research direction leads to the development of advanced technologies for health-related prediction, including facial expression recognition and human computing.

\end{IEEEbiography}
\begin{IEEEbiography}[{\includegraphics[width=1in,height=1.25in,clip,keepaspectratio]{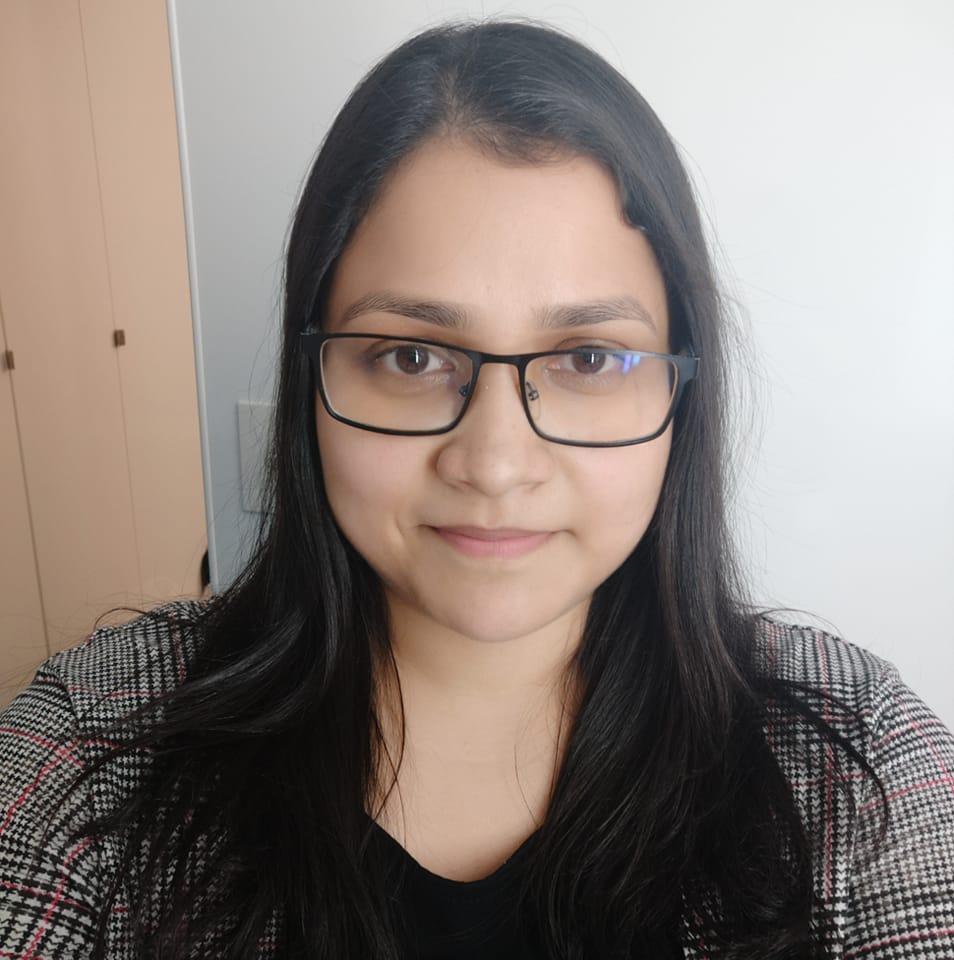}}]{Shreya Ghosh}
received a BTech degree in CSE from the Govt. College of Engineering and Textile Technology, India, and the MS(R) degree in computer science and engineering from the Indian Institute of Technology Ropar, India, and the PhD degree from Monash University, Australia, in 2022. She is currently a research academic at Curtin University. Her research interests include affective computing and computer vision.
 
\end{IEEEbiography}
\begin{IEEEbiography}[{\includegraphics[width=1in,height=1.25in,clip,keepaspectratio]{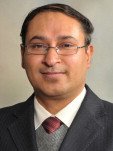}}]{Aneesh Krishna} is currently a Professor and Discipline Lead of Computing within the School of Electrical Engineering, Computing and Mathematical Sciences, Curtin University, Australia. He holds a PhD in computer science from the University of Wollongong, Australia. His research interests include AI for software engineering and model-driven development/evolution. His research is (or has been) funded by various Australian government agencies and companies.
\end{IEEEbiography}

\begin{IEEEbiography}[{\includegraphics[width=1in,height=1.25in,clip,keepaspectratio]{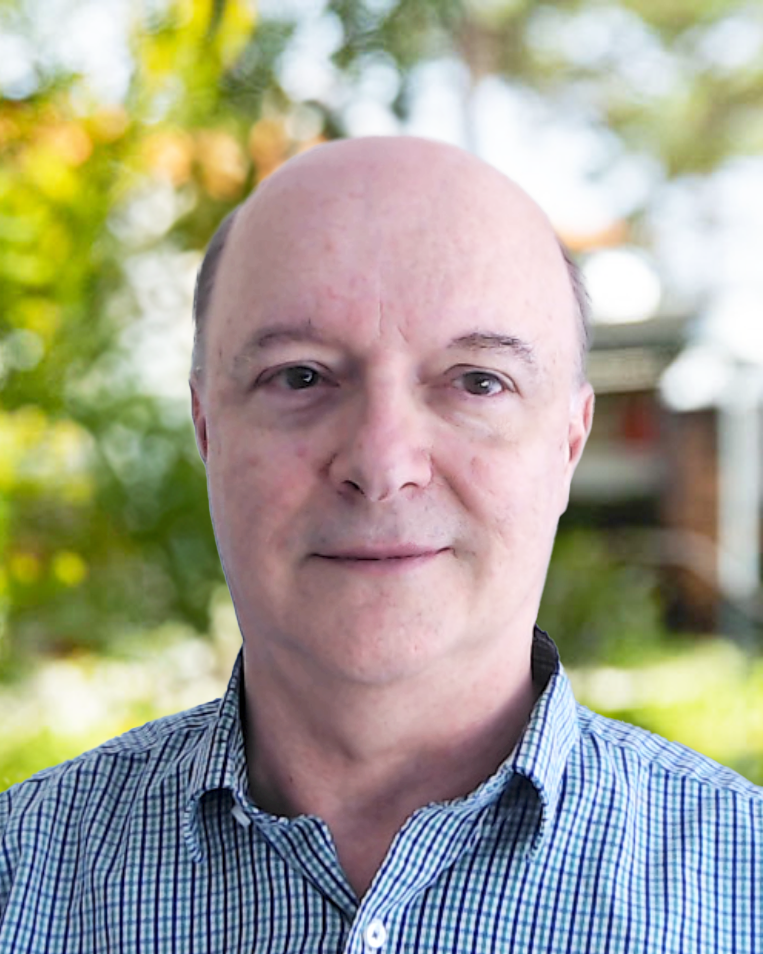}}]{Tom Gedeon}
is a Professor as the Human-Centric Advancements Chair in AI and was recently the Optus Chair in AI at Curtin University. Prior to this, he was a Professor and former Deputy Dean of the College of Engineering and Computer Science at ANU. He has over 400 publications and has run multiple international conferences. He is a former president of the Asia-Pacific Neural Network Assembly and former President of the Computing Research and Education Association of Australasia.
\end{IEEEbiography}







\end{document}